\documentclass[journal]{IEEEtran}
\usepackage[utf8]{inputenc}
\usepackage[english]{babel}

\usepackage{cite}
\usepackage{amsmath}
\usepackage{xcolor}
\usepackage{xspace}
\usepackage{tikz}
\usepackage{enumitem}

\usepackage{multirow}
\usepackage[final]{microtype}
\usepackage{subcaption}
\usepackage{seqsplit}
\usepackage{pdfpages}

\newcommand{\Design}{RAPIDx\xspace}

\definecolor{darkgreen}{RGB}{0,176,80}

\DeclareRobustCommand\encircle[1]{\tikz[baseline=(char.base)]{\node[shape=circle,fill,inner sep=0.75pt] (char) {\textcolor{white}{#1}}}}

\setlist[itemize]{noitemsep, topsep=0pt, leftmargin=*}

\usepackage{caption} 
\usepackage{titlesec}
\titlespacing*{\section}
{0pt}{1.0ex plus 0.6ex minus .2ex}{.4ex plus .1ex}
\titlespacing*{\subsection}
{0pt}{.3ex plus .1ex minus .0ex}{.1ex plus .0ex}
\titlespacing*{\subsubsection}
{0pt}{0ex plus 0ex minus .0ex}{0.ex plus .0ex}

\begin{document}

% \includepdf[pages=-]{resp_new.pdf}

\title{\Design: High-performance ReRAM Processing in-Memory Accelerator for Sequence Alignment}


\author{Weihong~Xu,
        Saransh~Gupta,
        Niema~Moshiri, 
        and~Tajana~Rosing,~\IEEEmembership{Fellow,~IEEE}% <-this % stops a space
\thanks{W.~Xu, N.~Moshiri, and~T.~Rosing are with University of California, San Diego (UCSD), La Jolla, CA 92093 USA (e-mail: wexu@ucsd.edu; niema@ucsd.edu; tajana@eng.ucsd.edu).}% <-this % stops a space
\thanks{Saransh~Gupta is now with IBM Research (e-mail: saransh@ibm.com).}% <-this % stops a space
\thanks{Part of this work was presented in International Symposium on Low Power Electronics and Design (ISLPED), 2019.}
% \thanks{Manuscript received April 19, 2005; revised August 26, 2015.}
\vspace{-0.5cm}
}


\markboth{IEEE TRANSACTIONS Trans ON COMPUTER-AIDED DESIGN OF INTEGRATED CIRCUITS AND SYSTEMS}
%,~Vol.~xx, No.~x, xx~xxxx}%
{Xu \MakeLowercase{\textit{et al.}}: \Design: High-performance ReRAM Processing in-Memory Accelerator for Sequence Alignment}




\maketitle

\begin{abstract}
Genome sequence alignment is the core of many biological applications. The advancement of sequencing technologies produces a tremendous amount of data, making sequence alignment a critical bottleneck in bioinformatics analysis. The existing hardware accelerators for alignment suffer from limited on-chip memory, costly data movement, and poorly optimized alignment algorithms. They cannot afford to concurrently process the massive amount of data generated by sequencing machines. In this paper, we propose a ReRAM-based accelerator, RAPIDx, using processing in-memory (PIM) for sequence alignment. RAPIDx achieves superior efficiency and performance via software-hardware co-design. First, we propose an adaptive banded parallelism alignment algorithm suitable for PIM architecture. Compared to the original dynamic programming-based alignment, the proposed algorithm significantly reduces the required complexity, data bit width, and memory footprint at the cost of negligible accuracy degradation. Then we propose the efficient PIM architecture that implements the proposed algorithm. The data flow in RAPIDx achieves four-level parallelism and we design an in-situ alignment computation flow in ReRAM, delivering $5.5$-$9.7\times$ efficiency and throughput improvements compared to our previous PIM design, RAPID. The proposed RAPIDx is reconfigurable to serve as a co-processor integrated into the existing genome analysis pipeline to boost sequence alignment or edit distance calculation. On short-read alignment, RAPIDx delivers $131.1\times$ and $46.8\times$ throughput improvements over state-of-the-art CPU and GPU libraries, respectively. As compared to ASIC accelerators for long-read alignment, the performance of RAPIDx is $1.8$-$2.9\times$ higher.
\end{abstract}


\IEEEpeerreviewmaketitle

\section{Introduction}
Genome techniques are becoming increasingly crucial in various fields. Modern genome analysis techniques have been applied to human DNA to diagnose genetic diseases by identifying disease-associated structural variants~\cite{seqoverview}. The genome sequence information is also used to infer the evolutionary history of an organism over time \cite{beast}. These sequences can also be analyzed to provide information on populations of viruses within individuals, allowing for a comprehensive understanding of underlying viral selection pressures \cite{minoritydetection}.

DNA sequence alignment is a key step in genome analysis that gains increasing significance due to the following reasons. First, several types of sequencing errors occur when the sequencing machine reads the genome. Additionally, genetic mutations and variations also introduce sequence differences. DNA alignment algorithms, like Needleman–Wunsch (NW) \cite{nw} and Smith-Waterman (SW) \cite{smith1981identification}, are used to identify the optimal match between the query and reference sequences. 
The other reason is that the alignment step has become the bottleneck of genome analysis pipeline because sequence alignment is a computation-intensive and memory-intensive workload, taking up 60-80\% runtime of popular genome analysis tools \cite{vsovsic2017edlib,minimap2,gasal2,bwamem}. Therefore, boosting DNA sequence alignment plays an important role in accelerating genome analysis.

Various algorithm optimizations have been developed for software libraries \cite{vsovsic2017edlib,minimap2,bwamem,langmead2012fast}. However, the limited computing resources of CPU severely restrict the achievable performance. These works fail to generate satisfactory processing throughput and energy efficiency. To this end, many efforts have been made to design acceleration solutions on ASIC \cite{liao2018adaptively,genasm,turakhia2018darwin}, GPU \cite{gasal2,de2016cudalign}, or FPGA \cite{arram2017leveraging} platforms. Through optimizing algorithm and hardware architecture, these accelerators have shown significant improvements in terms of efficiency and processing speed. However, the memory-intensive nature of DNA alignment algorithms makes them suffer from the limited on-chip as well as expensive data movement between off-chip memory and processing cores, incurring energy overhead caused by data movement.

\begin{figure}[t]
	\centering
	\begin{subfigure}{.5\textwidth}
		\centering
		\includegraphics[height=.4\linewidth]{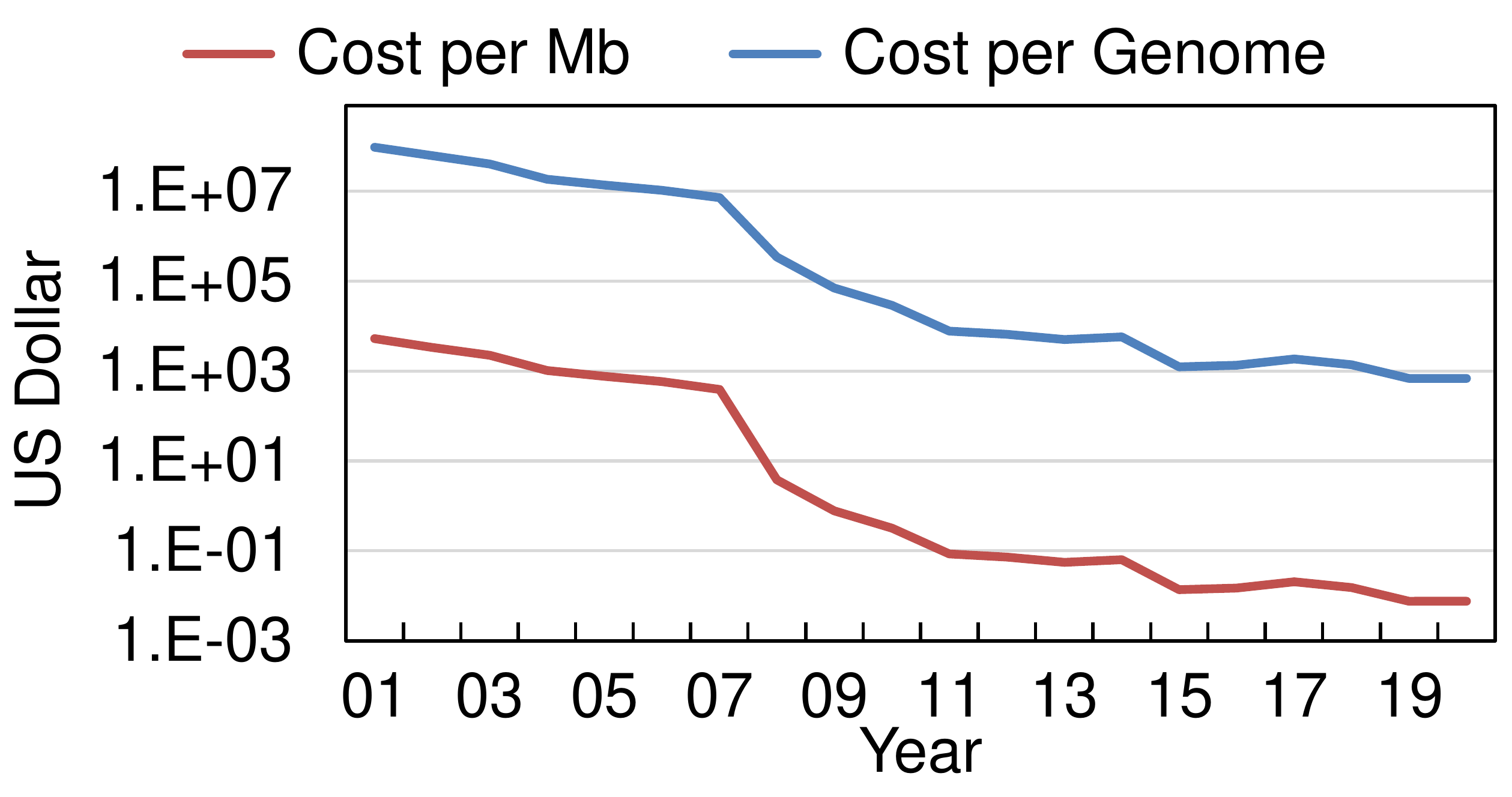}
	\end{subfigure}%
	\hfill
	\begin{subfigure}{.5\textwidth}
		\centering
		\includegraphics[height=.4\linewidth]{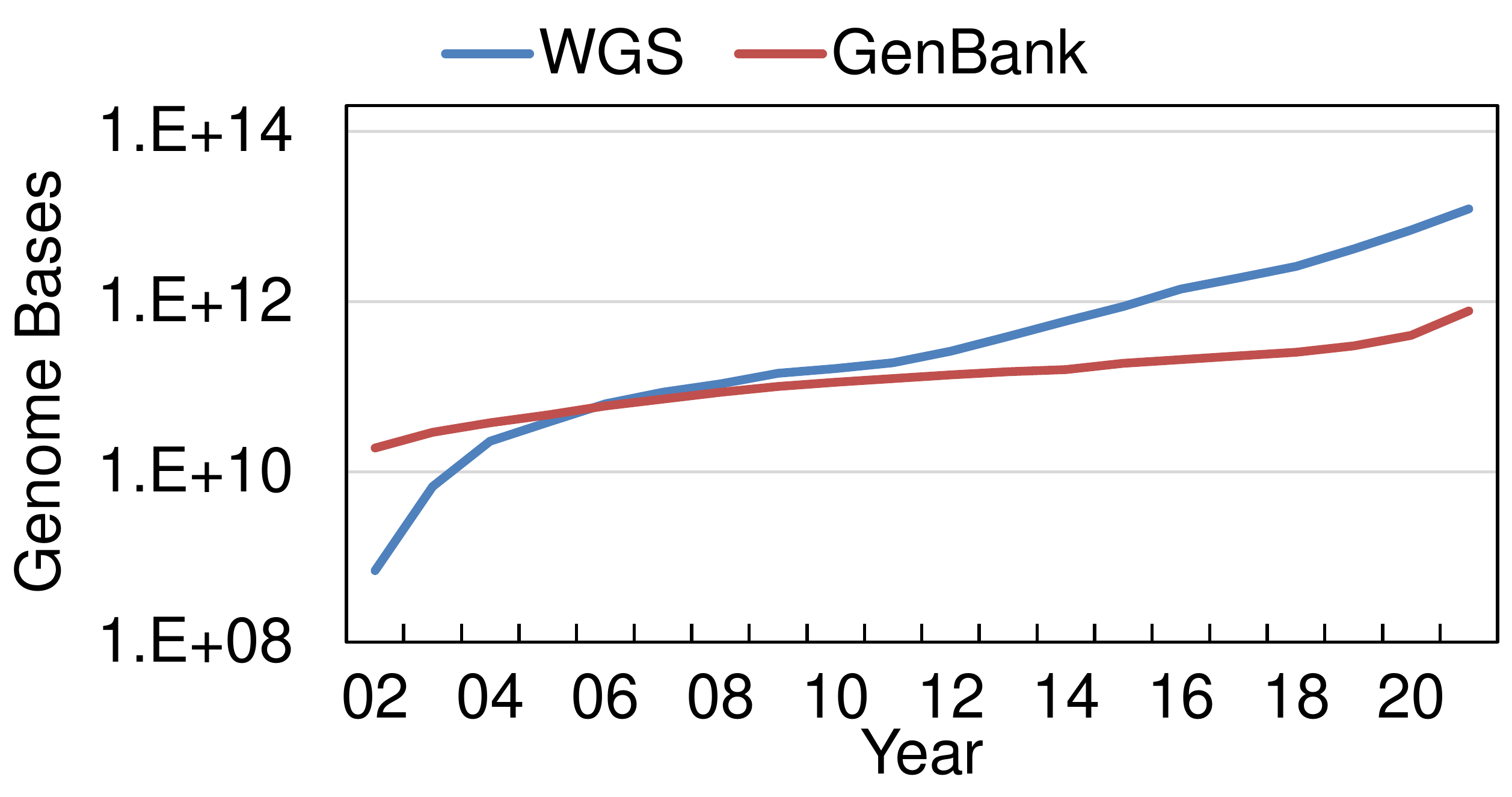}
	\end{subfigure}
	\caption{Trend of unit sequencing cost \cite{genome_cost} and genome data volume \cite{genbank} over the past decade.}
	\label{fig:gen_trend}
\end{figure}

The advent of high-throughput next-generation sequencing (NGS) technique \cite{wenger2019accurate} enlarges the gap between the processing capabilities of existing alignment accelerators and the rapidly generated genome data. 
Fig. \ref{fig:gen_trend} shows the unit cost of genome sequencing has plunged by over $10^4\times$ during the last decade. Meanwhile, the genome data volume of whole genome sequencing (WGS) and GenBank \cite{genbank} have also expanded by $10^2\times$ to $10^4\times$. The genome data growth has significantly surpassed Moore's Law, meaning that acceleration solutions with orders of magnitude higher efficiency are needed for sequence alignment. 
Processing in-memory (PIM) is promising to mitigate the data movement issue and provides massive parallelism. This is because PIM enables in-situ data computation inside memory, thereby throttling the latency and energy of data movement~\cite{gokhale1995processing, ahn2015pim, li2016pinatubo, gupta2019nnpim}. Existing PIM-based accelerators for genome analysis \cite{kaplan2017resistive,bioseal,huangfu2018radar,angizi2019aligns,zokaee2018aligner,gupta2019rapid} take PIM's advantages of high data parallelism and low-cost data movement, showing orders of magnitude efficiency and performance improvements over CPU and GPU.

We previously presented the PIM architecture for sequence alignment, called RAPID \cite{gupta2019rapid}, which computes DNA alignment in memory. However, RAPID has the following deficits. First, the original DP algorithm \cite{nw} used by RAPID is sub-optimal since it is unable to measure the affine gap penalty, which has been widely used in software libraries \cite{minimap2,gasal2} and shown optimal alignment quality \cite{liu2008barking}. Second, RAPID does not consider software-hardware co-optimization, thereby wasting a large amount of energy and computing resources on redundant computations. Recent works \cite{bwamem,seedex} demonstrate DP alignment algorithm exhibits great redundancy, and most of computation can be skipped using banded alignment \cite{chao1992aligning} to accelerate the alignment process at the cost of negligible accuracy degradation. 
In this paper, we propose a software-hardware co-design, \Design, that exploits digital PIM techniques on ReRAM to enable a highly parallel and more energy-efficient acceleration for sequence alignment. The key contributions of this work can be summarized as follows:
\begin{itemize}[leftmargin=*]
	\item \textbf{PIM-friendly dynamic programming (DP) alignment}: We consider the affine gap penalty to construct more accurate scoring functions. Then we propose the adaptive banded parallelized DP alignment that is friendly for PIM implementation. The proposed alignment algorithm reduces the required arithmetic precision from 32-bit to only 5-bit and obtains higher data parallelism. Meanwhile, the adaptive wavefront direction and bandwidth schemes significantly reduce memory footprint and computational complexity by over $10\times$ at the cost of $<0.15\%$ accuracy loss.
	
	\item \textbf{High-performance PIM architecture}: We propose efficient PIM architecture for \Design, which achieves four-level data parallelism. \Design leverages in-situ PIM operations \cite{gupta2018felix} to perform low-energy and row-parallel in-memory alignment. Our peripheral circuits implement fast traceback as well as complex functions not friendly for PIM. Compared to previous RAPID \cite{gupta2019rapid}, RAPIDx shows $5.5\times$ latency reduction and $6.2\times$ energy improvements.
	
	\item \textbf{System optimization and reconfigurable design}: We design novel PIM computing operations that are reconfigurable to support multiple types of alignment scoring as well as edit distance computation. This makes \Design a multi-purpose accelerator that is flexible to support alignment and edit distance computations. We also analyze several possible limiting factors when integrating \Design into existing computing system, including ReRAM cell's limited endurance, switching speed, and system considerations. 
	
	%show that \Design .
	
	\item \textbf{Improvements and accelerations}: We compare \Design with state-of-the-art CPU baselines (Minimap2~\cite{minimap2} and Edlib~\cite{vsovsic2017edlib}), GPU baseline (GASAL2~\cite{gasal2}), and ASIC baselines (ABSW~\cite{liao2018adaptively} and GenASM~\cite{genasm}) on various workloads. For short-read alignment, \Design delivers an average $131.1\times$ and $46.8\times$ higher throughput compared to Minimap2 \cite{minimap2} and GASAL2 \cite{gasal2}, respectively. For long-read alignment, $1.8\times$ to $2.9\times$ throughput improvements are observed over ABSW~\cite{liao2018adaptively} and GenASM~\cite{genasm}. For edit distance calculation, \Design obtains up to $321\times$ speedup over Edlib \cite{vsovsic2017edlib}.
\end{itemize}

\begin{figure*}[t]
	\centering
    \includegraphics[width=\linewidth]{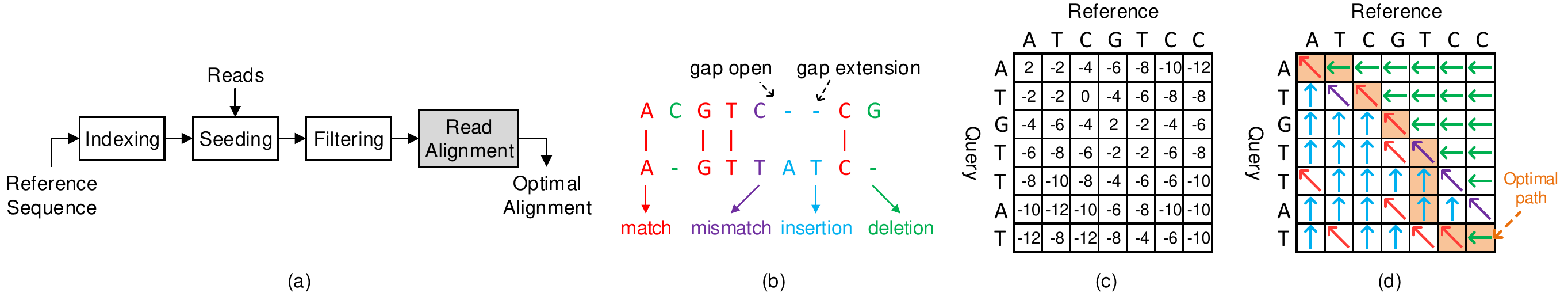}
	\caption{(a) The pipeline of genome sequence analysis. (b) Alignment example of sequences $\mathsf{ACGTCCG}$ and $\mathsf{AGTTATC}$ with affine gap penalties, (c) Score matrix, (d) Traceback matrix.}
    \vspace{-0.4cm}
	\label{fig:dna_genome_flow}
\end{figure*}

%%%%%%%%%%%%%%%%%%%%%%%%%%%%%%%%%%%%%%%%%%%%%%%%%%%%%%%%%%%%%%
\section{Related Work}

\subsection{Software for Sequence Alignment}
Several software libraries \cite{vsovsic2017edlib,minimap2,gasal2,bwamem} have been developed for boosted genome analysis. The main point is optimizing the SW algorithm and CPU/GPU datapath to deliver accurate and fast sequence alignment. BWA-MEM \cite{bwamem} is software to map DNA sequences against large reference genomes. BWA-MEM aligns the given sequences using Burrows-Wheeler Transform (BWT) \cite{bwt}. However, the memory footprint of aligning long genome is large and the irregular memory access of BWT limits the processing speed. Edlib \cite{vsovsic2017edlib} is a C++ library that exploits Myers’s bit-vector algorithm \cite{myers1999fast} to parallelize the SW-based alignment. To realize more accurate and efficient alignment, Minimap2 \cite{minimap2} introduces two promising optimization strategies, banded alignment \cite{chao1992aligning} and difference-based SW \cite{suzuki2018introducing}, which can be fitted into the datapath of single instruction, multiple data (SIMD). Minimap2 generates over $10\times$ speedup over BWA-MEM. Even though these software libraries achieve fine-grain optimization, the limited computing resources on CPU fail to provide opportunities for further acceleration. Some researchers shift the focus to GPU-based acceleration. CUDAlign 4.0 \cite{de2016cudalign} increases the parallelism by splitting each SW alignment into multiple GPUs and reducing the data dependency of the traceback process. GASAL2 \cite{gasal2} optimizes the data organization and develops efficient kernels for multiple sequence alignment workloads.  These libraries exploit the abundant computing resources on GPU. But the resulted efficiency is not high because optimizations for SW algorithms are lacked due to the architectural limitations of GPU. In this work, \Design is a software and hardware co-design that realizes algorithm and hardware optimizations at the same time.

\subsection{Hardware Acceleration for Sequence Alignment}
\textbf{ASIC Accelerator}:
Various hardware accelerators \cite{turakhia2018darwin,liao2018adaptively,genasm,bioseal,kaplan2017resistive,huangfu2018radar,zokaee2018aligner,gupta2018felix} have be presented to obtain higher energy efficiency and speedup for genome analysis. For ASIC designs, one challenge is how to realize long-read alignment under the constraints of limited on-chip memory. Darwin \cite{turakhia2018darwin} proposes near-optimal tiling methods to align arbitrary sequence lengths, only requiring constant memory space. ABSW \cite{liao2018adaptively} leverages the tiling schemes \cite{turakhia2018darwin} and implements an adaptively banded alignment on ASIC, achieving significant efficiency improvement. GenASM \cite{genasm} proposes an approximate string matching algorithm and a systolic-array-based accelerator to increase data parallelism while reducing memory footprint. Although prior works employ a variety of optimizations, the limited on-chip memory is still the bottleneck when aligning long sequences.

\noindent
\textbf{PIM Accelerator}:
PIM is a promising solution to increase data parallelism and energy efficiency via computing data in situ \cite{zokaee2018aligner, angizi2019aligns, angizi2020pim, li2016pinatubo}. The PIM-based alignment designs proposed in PRINS \cite{kaplan2017resistive} and BioSEAL~\cite{bioseal} accelerate algorithms using resistive content addressable memory (CAM). But the sequential associative search incurs a large amount of write operation and internal data movement, degrading efficiency, lifetime, and storage efficiency. Another set of works accelerates short read alignment, where long sequences are broken down into smaller sequences and heuristic methods are applied. AlignS \cite{angizi2019aligns}, AligneR \cite{zokaee2018aligner} and PIM-Aligner \cite{angizi2020pim} exploit FM-index algorithm and PIM to realize short-read alignment. However, FM-index incurs irregular memory access, and is hard to exploit the data parallelism of PIM. RAPID \cite{gupta2019rapid} is a ReRAM-based PIM accelerator to implement in-situ alignment computation in the memory, which drastically reduces the data movement. However, the adopted algorithm in RAPID is sub-optimal and requires quadratic complexity, limiting its capability of aligning long sequences. In this work, we present several optimizations for alignment algorithms and hardware architecture to fully leverage the highly parallel PIM while providing satisfactory alignment quality. Our design, \Design, delivers up to $9.3\times$ alignment efficiency improvement compared to other PIM baselines.

\section{Background}
\subsection{Genome Sequence Analysis}
\subsubsection{Overall Pipeline}
A typical pipeline of modern genome sequencing analysis \cite{minimap2,bwamem,langmead2012fast} involves indexing, seeding, filtering, and read alignment steps as shown in Fig. \ref{fig:dna_genome_flow}~(a). For the indexing phase, the entire reference sequence is stored into special data structures, like BWT \cite{bwt} and FM-indexing. The indexing is for quickly obtaining the location of query sequence in the reference sequence. Then, the seeding process uses the indexing information to query the potential mapping locations of genome reads. The filtering step discards invalid candidates or combines nearby candidates from the seeding step. Finally, the genome reads are aligned against the reference sequence around the candidate location using the SW algorithm. Among these steps, the most time-consuming step is read alignment used to determine how the read sequence can be optimally mapped to the reference sequence.

\subsubsection{Sequence Alignment with Affine Gap Penalty}
The sequence alignment can be described as finding the maximum alignment score between the reference sequence $R = r_1, r_2, ..., r_m$ and the query sequence $Q = q_1, q_2, ..., q_n$. Natural evolution and mutation as well as experimental errors during sequencing poses two types of changes in sequences - substitutions and indels. A substitution changes a base of the sequence with another, leading to a mismatch whereas an indel either inserts or deletes a base. Fig. \ref{fig:dna_genome_flow}~(b) shows the comparison of two sequences, $R = \mathsf{ACGTCCG}$ and $Q= \mathsf{AGTTATC}$. The left part rigidly compares the $i\text{th}$ base of $Q$ with $R$, where \textcolor{red}{match} and \textcolor{violet}{mismatch} are considered. The right part assumes a different alignment that involves \textcolor{cyan}{insertion} and \textcolor{darkgreen}{deletion}. Note that the notation of dashes ($-$) is conceptual, and are used to illustrate a potential scenario that one sequence has been (or can be) evolved to the other.

Most sequence alignments are categorized into global or local alignment. The global and local alignments can be optimally addressed by NW algorithm \cite{nw} and SW algorithm \cite{smith1981identification}, respectively. NW and SW both build up and compute the optimal alignment sequence based on DP \cite{altschul1990basic, lipman1985rapid}. DP-based methods involve forming alignment matrices, which are used to compute scores of various alignments based on a pre-defined scoring function. The scoring function is essential for accurate alignment since it is used to update the scoring matrix in DP. The previous work \cite{banerjee2019asap} mostly uses the scoring function with linear gap penalty, where the penalty is increasing linearly with the gap length. However, the linear gap penalty is insufficient to accurately evaluate the alignment scores for those sequences with the same total gap length. The gap-less sequence is normally more biologically meaningful compared to the sequence with more gaps. In this work, we adopt the scoring function with affine gap penalties \cite{gotoh1982improved} that consider the number and length of gaps. Fig. \ref{fig:dna_genome_flow} shows an example of alignment between sequence $R = \mathsf{ACGTCCG}$ and $Q= \mathsf{AGTTATC}$ using affine gap penalties. The updating rules for scoring matrices in DP with affine gap penalty can be expressed as:
\begin{align}\label{eq:align_affine}
    \small
    \begin{split}
        E_{i, j} &= \max \begin{cases}
            H_{i-1, j} - o\\
            E_{i-1, j} - e
        \end{cases} 
        \hfill
        F_{i, j} = \max\begin{cases}
            H_{i, j-1} - o\\
            F_{i, j-1} - e
        \end{cases} 
        \\
        H_{i, j} &= \max\{E_{i, j},F_{i, j},H_{i-1, j-1} - s(r_j, q_i)\}
    \end{split}
\end{align}
where $E$ and $F$ denote the alignment matrices that store the scores of insertion and deletion, respectively. $H$ is the alignment score matrix that stores the total scores. $s(r_j, q_i)$ denotes the score of match $A$ or mismatch $B$ by comparing $r_j$ and $q_i$. The gap opening penalty is $o$ while $e$ denotes the gap extension penalty. Fig. \ref{fig:dna_genome_flow}~(c) shows an example of score matrix $H$ calculated using Eq (\ref{eq:align_affine}) with penalties $A=2, B=4, o=4, e=2$. A traceback phase is required to construct the optimal alignment path after the computation for all alignment matrices. The traceback matrix in Fig. \ref{fig:dna_genome_flow}~(d) stores the path information. For global alignment, the traceback starts from the cell at the bottom-right corner while local alignment starts from the cell with the maximum score.

\subsection{Difference-based Dynamic Programming (DP) Alignment}\label{subsec:diff_align}
The updating function in Eq. (\ref{eq:align_affine}) has the following limitations. The maximum value in the alignment matrix scales up linearly with the matrix dimension. The data bit width needs to be increased as the sequence length increases to avoid computation overflow. Previous accelerations \cite{liao2018adaptively,turakhia2018darwin} use a fixed bit width in the worst case, resulting in low computation efficiency. To resolve this issue, the original DP updating is rewritten into a computation-efficient form, named the difference-based formulation \cite{suzuki2018introducing}. The basic idea is to store and compute the value difference of adjacent elements instead of the full-precision value in the alignment matrix, thus reducing the required arithmetic precision. As shown in the left side of Eq. (\ref{eq:align_diff}), four matrices $\Delta H$, $\Delta V$, $\Delta E$, and $\Delta F$ are used to store the difference values. After substituting the four difference matrices into Eq. (\ref{eq:align_affine}), the alignment matrices ($H$, $E$, and $F$) are converted into the following difference-based formulation:
\begin{align}\label{eq:align_diff}
    \small
    \begin{split}
        &\begin{cases}
            \Delta H_{i,j} = H_{i,j} - H_{i-1,j} \\ 
            \Delta V_{i,j} = H_{i,j} - H_{i,j-1} \\ 
            \Delta E_{i,j} = E_{i+1,j} - H_{i,j} \\ 
            \Delta F_{i,j} = F_{i,j+1} - H_{i,j}
        \end{cases} 
        \\
        \Rightarrow &
        \begin{cases} 
            A_{i,j} = \max\begin{cases}
                s(i,j), \\
                \Delta E_{i-1,j} + \Delta V_{i-1,j}, \\
                \Delta F_{i,j-1} + \Delta H_{i,j-1}
            \end{cases}  \\ 
            \Delta H_{i,j} = A_{i,j} - \Delta V_{i-1,j} \\ 
            \Delta V_{i,j} = A_{i,j} - \Delta H_{i,j-1} \\ 
            \Delta E_{i,j} = \max \{ -o, \Delta E_{i-1,j} - \Delta H_{i,j}\} - e \\ 
            \Delta F_{i,j} = \max \{ -o, \Delta F_{i,j-1} - \Delta V_{i,j} \} - e
        \end{cases}
    \end{split}
\end{align}
where an intermediate variable $A_{i, j}$ is added to the computation. It should be noted that Eq. (\ref{eq:align_diff}) only changes the expression of original DP in Eq. (\ref{eq:align_affine}) while retaining the identical information. Eq. (\ref{eq:align_diff}) can generate the identical alignment results as Eq. (\ref{eq:align_affine}).

There are two benefits of the difference-based alignment in Eq. (\ref{eq:align_diff}). First, the arithmetic precision requirement is significantly reduced. According to \cite{minimap2, suzuki2018introducing}, the data range of $\Delta H_{i,j}$ and $\Delta V_{i,j}$ are bounded by $[-o-e, -e]$ while $\Delta E_{i,j}$ and $\Delta F_{i,j}$ are bounded by $[-o-e, M+o+e]$, where $M$ denotes the maximum value of $s(i,j)$. Compared to the full-precision alignment, the difference-based representations only needs $\lceil \log_{2}(M+2o+2e+1)\rceil$-bit integer to calculate the alignment. Second, the required data precision is only determined by the used affine gap scores while independent with the sequence length. This property allows us to use a unified data bit width for different sequence lengths. For example, we use 5-bit integer for computing alignment and 3-bit integer for calculating edit distance as introduced in Section \ref{subsec:dyn_precision}.

\subsection{Digital Processing In-Memory (PIM)}\label{sec:pim_back}

Various types of memory devices are used for PIM to resolve the ``memory wall'' problem, including MRAM \cite{angizi2020pim,angizi2019aligns}, PCM, and SRAM \cite{lee2020bit}. MRAM suffers from severe read disturbance when the memory density increases \cite{boukhobza2017emerging}. 
ReRAM has higher memory density than MRAM and SRAM because the ReRAM cell is much smaller than MRAM and SRAM. Moreover, ReRAM has lower leakage power compared to other devices, making it an energy-efficient candidate for PIM. FeFET \cite{reis2018computing} and NAND flash \cite{kim2021embedded} are the other two potential PIM candidates that are still in early development phase while ReRAM has been physically verified at scale \cite{xue202116}. 
ReRAM has higher error rates, but this is not a significant issue for alignment as alignment algorithms are already statistical in nature, and can tolerate significant errors at bit level.
Considering all these benefits, we choose ReRAM-based PIM in this work.

\begin{figure}[t]
	\centering
	{\includegraphics[width=0.7\columnwidth]{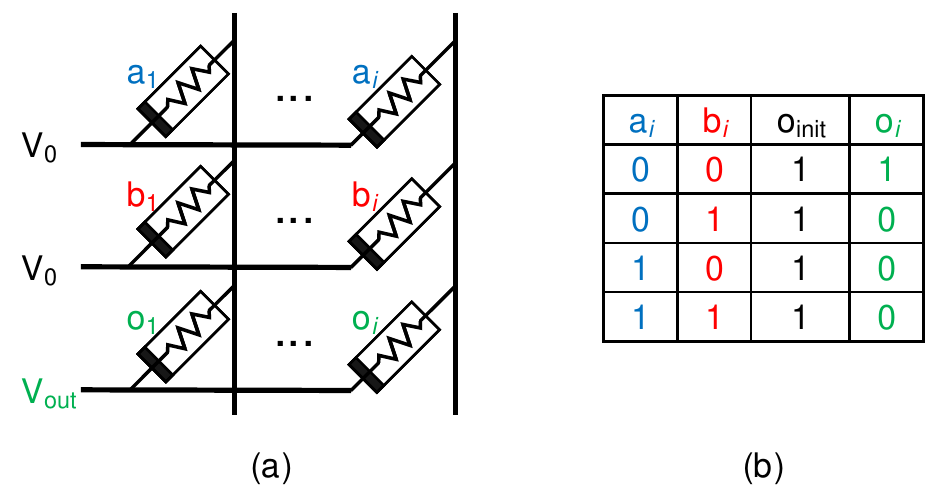}} 
	\caption{Implementing NOR operation using ReRAM-based digital processing in-memory (PIM).}
	\label{fig:background}
\end{figure}

Traditionally, PIM with memristors is based on reading currents through different cells. However, some recent work has demonstrated ways, both in literature~\cite{gupta2018felix, talati2016logic, borghetti2010memristive} and by fabricating chips~\cite{jang2018memristive}, to implement logic using memristor switching. Digital PIM exploits variable switching of memristors. The output device switches whenever the voltage across it exceeds a threshold~\cite{kvatinsky2015vteam}. This property can be exploited to implement a variety of logic functions inside memory~\cite{gupta2018felix, talati2016logic}. 
Fig. \ref{fig:background} shows an example of implementing NOR operation using ReRAM-based PIM \cite{gupta2018felix}. A voltage $V_{0}$ is in parallel applied to the rows that contain the operand cells $a_i$ and $b_i$. The output cell $o_i$ switches to low voltage status (logical `0') from initial logical `1' whenever one or more inputs are `1's, resulting in logical \texttt{NOR} operation. 
Since \texttt{NOR} is a functionally complete logic gate, it can be used to implement other logic operations like addition~\cite{talati2016logic} and multiplication~\cite{haj2018efficient}. For example, 1-bit addition (inputs being $A, B, C$) can be represented in the form of \texttt{NOR} as:
\begin{align}\label{eq_sum}
    \small
    \begin{split}
        C_{out} = ((A+B)'+(B+C)'+(C+A)')'\\
        S = (((A'+B'+C')'+((A+B+C)'+C_{out})')')'
    \end{split}
\end{align}
where $C_{out}$ and $S$ are the generated carry and sum bits of addition. $(A+B+C)'$, $(A+B)'$, and $A'$ represent $NOR(A,B,C)$, $NOR(A,B)$, and $NOR(A,A)$, respectively.

In-memory operations are in general slower than the corresponding CMOS-based implementations because memristor devices switch slowly. However, PIM architectures can provide significant speedup when it is exposed massive parallelism. Meanwhile, the long processing latency is amortized due to the high parallelism. In this work, \Design utilizes two types of PIM operations (XOR and addition) introduced in FELIX \cite{gupta2018felix} to perform in-memory alignment computation. This is because FELIX's PIM primitives achieve the same or significantly better latency, memory consumption, and efficiency than other digital PIM schemes~\cite{kvatinsky2014magic,talati2016logic}. The other digital PIM scheme~\cite{imani2019floatpim} for floating-point arithmetic is not suitable for the fixed-point arithmetic in \Design. 

Specifically, the XOR and 1-bit addition are realized through:
\begin{itemize}
	\item \textbf{XOR:}
	XOR ($\oplus$) can be expressed by OR ($+$), AND (.), and NAND ($(.)'$) as $A\oplus B = (A+B).(A.B)'$. We first calculate OR and then use its output cell to implement NAND. This operation is executed in parallel over all the columns of two rows. This logic just requires 2 cycles and one additional memristor device, which acts as the output cell.
	\item \textbf{Addition:}
	Let A, B, and $C_{in}$ be 1-bit inputs of addition, and S and $C_{out}$ the generated sum and carry bits respectively. Then, S is implemented as two serial in-memory XOR operations (A$\oplus$B)$\oplus$C. $C_{out}$, on the other hand, can be executed by inverting the output of the Min function proposed in \cite{gupta2018felix}. Addition takes a total of 6 cycles and similar to XOR, we parallelize it over all columns in two rows.
\end{itemize}

%%%%%%%%%%%%%%%%%%%%%%%%%%%%%%%%%%%%%%%%%%%%%%%%%%%%%%%%%%%%%%%%%%%%%%%%%%%%%%%%%%%%%%%%
%%%%%%%%%%%%%%%%%%%%%%%%%%%%%%%%%%%%%%%%%%%%%%%%%%%%%%%%%%%%%%%%%%%%%%%%%%%%%%%%%%%%%%%%

\begin{figure*}[t]
	\centering
	\includegraphics[width=0.75\linewidth]{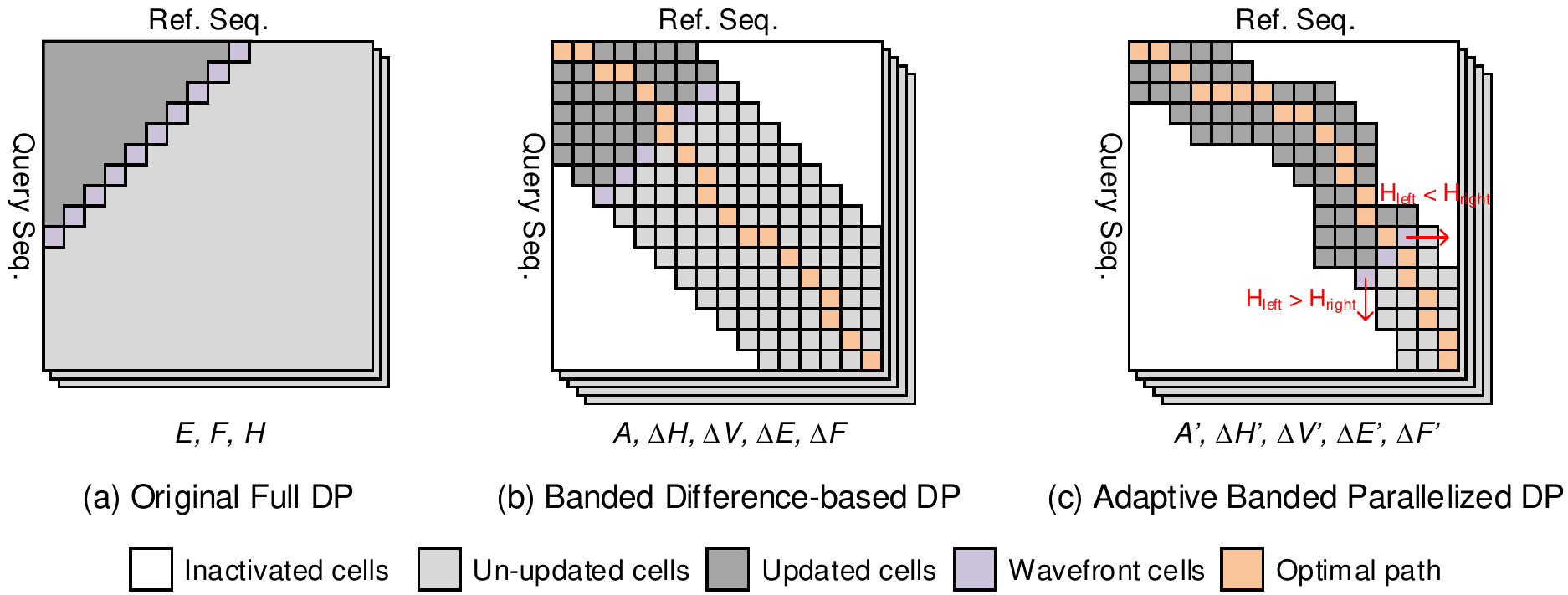}
	\caption{Illustration of three variants of DP alignment algorithms. Bandwidth $B=6$ in (b) and $B=3$ in (c).}
    \vspace{-0.2cm}
	\label{fig:adaptive_band}
\end{figure*}

\section{Efficient Alignment in \Design} \label{sec:diff_align_alg}
In this section, we first analyze the challenges of realizing efficient in-memory alignment using digital PIM. Then we propose the adaptive banded parallelized DP alignment to balance performance and accuracy loss.

\subsection{Challenges of Alignment using PIM}
\subsubsection{Data Bit Width and Latency}
Compared to CMOS-based circuits, the slow switching speed of ReRAM cells incurs long latency when implementing PIM operations in Section \ref{sec:pim_back}. For example, 1-bit PIM addition takes 6 to 12 clock cycles \cite{gupta2018felix}. As discussed in Section \ref{subsec:diff_align}, the data bit width and range grow linearly with the sequence length. The previous accelerators \cite{gupta2019rapid,bioseal} adopt the original DP algorithm which uses 32-bit integers to guarantee lossless alignment. However, 32-bit integer is over-designed and incurs long processing latency when aligning short sequences ($<$1kbp) since the lower 12-bit width is enough to provide sufficient data dynamic range \cite{liao2018adaptively}. Therefore, developing an alignment algorithm using low bit-width data is beneficial to reduce PIM latency. The difference-based DP alignment in Section \ref{subsec:diff_align} is a potential solution to alleviate this as it needs fixed data width independent of sequence length.

\subsubsection{Data Parallelism}
ReRAM-based PIM architectures \cite{huangfu2018radar,gupta2019rapid,bioseal,angizi2020pim} offer substantial opportunities of extending the data parallelism. High parallelism amortizes the incurred long latency of PIM operations. One example is the row-parallel PIM operation \cite{gupta2018felix,bioseal}, where the bit-serial computation can be performed in the entire memory row simultaneously. How to exploit the architectural parallelism of ReRAM is key to attaining high alignment throughput. The other challenge from the algorithm is how to expose enough parallelism to ReRAM. For DP alignment, adjacent cells in alignment matrices exhibit data dependency. Previous works \cite{turakhia2018darwin,minimap2, liao2018adaptively, suzuki2018introducing,bioseal} utilize the wavefront parallelism based on the fact that cells over anti-diagonal have no data dependency. Unfortunately, this parallelism is far enough for PIM architecture.

\subsubsection{Complexity and Accuracy}
Fig. \ref{fig:adaptive_band} (a) illustrates the full DP alignment using Eq. (\ref{eq:align_affine}), where all cells in the matrices with shape $m\times n$ need to be computed ($m$ and $n$ denote the lengths of reference and query sequences, respectively). The complexity is prohibitive when aligning long sequences. Banded alignment \cite{chao1992aligning,seedex} is an effective method to reduce the complexity from quadratic to near-linear. It should be noted that the banded approach is an approximate algorithm that may introduce accuracy degradation. One simple solution is to use a fixed and wide bandwidth ($B=128$) as \cite{liao2018adaptively}. But this degrades the throughput and performance gain since wider bandwidth leads to higher complexity. The challenge is how to select narrow bandwidth for various lengths while ensuring the optimality of results.

\subsection{Adaptive Banded Parallelized DP Alignment}\label{subsec:adap_band_para_dp}
We propose the adaptive banded parallelized DP alignment to resolve the above-mentioned challenges. The difference-based alignment in Eq. (\ref{eq:align_diff}) relaxes the requirement of data precision and reduces the bit width for DP alignment. However, the computation of $\Delta H_{i,j}, \Delta V_{i,j}, \Delta E_{i,j}$, and $\Delta F_{i,j}$ can only be accomplished in a serial manner. Specifically, $A_{i,j}$ needs to be first computed before updating $\Delta H_{i,j}$ and $\Delta V_{i,j}$. Then the values of $\Delta V_{i,j}$ and $\Delta E_{i,j}$ require the newly updated $\Delta H_{i,j}$ and $\Delta V_{i,j}$. Consequently, parallelizing the computation for each updating step is difficult due to the inherent data dependency. We resolve this issue through further transforming Eq. (\ref{eq:align_diff}) into a parallelized version similar to \cite{suzuki2018introducing}. The variables in Eq. (\ref{eq:align_diff}) are rewritten as the top part of Eq. (\ref{eq:par_align_diff}), where auxiliary $o$ and $e$ values are added to each variable in Eq. (\ref{eq:align_diff}). After substituting it into Eq. (\ref{eq:align_diff}), we have the parallelized difference-based alignment as follows:
\begin{align}\label{eq:par_align_diff}
    \small
    \begin{split}
        &\begin{cases}
            A_{i,j}' = A_{i,j}+2o+2e\\
            \Delta H_{i,j}' = \Delta H_{i,j} + o + e \\ 
            \Delta V_{i,j}' = \Delta V_{i,j} + o + e \\ 
            \Delta E_{i,j}' = \Delta E_{i-1,j} + \Delta V_{i-1,j} + 2o + 2e \\ 
            \Delta F_{i,j}' = \Delta F_{i,j-1} + \Delta H_{i,j-1} + 2o + 2e
        \end{cases}
        \\
        \Rightarrow
        &\begin{cases} 
            A_{i,j}' = \max \{s'(i,j), \Delta E_{i-1,j}', \Delta F_{i,j-1}' \} \\ 
            \Delta H_{i,j}' = A_{i,j}' - \Delta V_{i-1,j}' \\ 
            \Delta V_{i,j}' = A_{i,j}' - \Delta H_{i,j-1}' \\ 
            \Delta E_{i,j}' = \max \{ A_{i,j}', \Delta E_{i-1,j}' + o \} - \Delta H_{i,j-1}' \\ 
            \Delta F_{i,j}' = \max \{ A_{i,j}', \Delta F_{i,j-1}' + o \} - \Delta V_{i-1,j}'
        \end{cases}
    \end{split}
\end{align}
where $\Delta H_{i,j}'$ and $\Delta V_{i,j}'$ only depend on new $A_{i,j}'$ and previous $\Delta V_{i-1,j}'$ and $\Delta H_{i,j-1}'$, respectively. Likewise, $\Delta E_{i,j}'$ and $\Delta F_{i,j}'$ can be calculated by the old $\Delta H_{i,j-1}'$ and $\Delta V_{i-1,j}'$ from the previous iteration. In this case, the relaxed data dependency between four alignment matrices provides higher computation parallelism. After obtaining $A_{i,j}'$, the computation of $\Delta H_{i,j}'$, $\Delta V_{i,j}'$, $\Delta E_{i,j}'$, and $\Delta F_{i,j}'$ can be conducted in parallel to shorten the processing latency. We call this the alignment matrix level parallelism. The data range of four alignment matrices is shifted to $[0, M+2o+2e]$ from $[-o-e, M+o+e]$, requiring the same bit width as Eq. (\ref{eq:align_diff}).

The banded alignment \cite{chao1992aligning} significantly reduces the complexity based on the observation that the optimal alignment path normally locates not far away from the diagonal of alignment matrices. The reduction is achieved by limiting the cells in alignment matrices that need to be computed. Fig. \ref{fig:adaptive_band} (b) shows the banded DP alignment that only computes the cells located within a bandwidth $B=6$ of the diagonal, whereas the rest cells are inactivated. In this way, only $B$ wavefront cells (the cells that are updated simultaneously) are computed and moved over the main diagonal in each iteration. Bandwidth and wavefront direction are the two key factors that determine the accuracy and efficiency of banded alignment. The adaptive banded parallelized alignment adopted by \Design is adaptive in the sense of bandwidth and wavefront direction as follows: 
\subsubsection{Adaptive Bandwidth}
A narrow bandwidth $B\ll m,n$ helps to perform a low-complexity alignment as the banded DP has $\mathcal{O}(mB)$ complexity. To balance the algorithm efficiency and accuracy, the bandwidth $B$ used in \Design is adaptive based on the processed sequence length. The other factor to consider when choosing the bandwidth is the inflexibility of ReRAM-based PIM. The proper bandwidth needs to be determined before alignment computation. 
To this end, we express the relationship between bandwidth $B$ and sequence length $L$ as $B = \min(w+0.01\times L, 100)$, where $w$ denotes the base bandwidth that determines the narrowest bandwidth while $B$ is set to the multiple of $w$. The function limits the maximum bandwidth to 100 because previous BWA-MEM \cite{bwamem} shows $B=100$ is enough to guarantee optimal alignment for all sequence lengths. On the other hand, a band with less than 20 is enough for over $99\%$ cases as demonstrated in \cite{seedex} but a too narrow band may not guarantee the optimality of alignment for long reads. This is because current long-read techniques (see Table \ref{table:error_profile}) incur much more errors and the narrow band can not fully cover the optimal path. Thus, we empirically select the $0.01$ coefficient to adaptively determine the minimum bandwidth that provides negligible degradation according to $L$. Based on the length of the given sequences, the bandwidth $B$ can be pre-determined before alignment. We provide detailed experiments in Section \ref{subsec:alg_valid} to guide the selection of the $0.01$ coefficient and the best $w$ that only introduce negligible accuracy loss.

\subsubsection{Adaptive Wavefront Direction}
The wavefront cells in Fig. \ref{fig:adaptive_band} (b) and (c) can move either rightward or downward in each iteration. The alignment tools, like Minimap2 \cite{minimap2} and BWA-MEM \cite{bwamem}, mostly use a pre-defined direction in Fig. \ref{fig:adaptive_band} (b), such that the wavefront moves towards the main diagonal. When we use narrow bandwidth ($B=3$) in Fig. \ref{fig:adaptive_band} (c), simply computing the wavefront over the main diagonal may not obtain the optimal results because the fixed wavefront direction lacks flexibility and is unable to cover the optimal path. To this end, we use a simple adaptive wavefront direction scheme to dynamically adjust the moving direction of wavefront cells as in Fig. \ref{fig:adaptive_band} (c). The direction is decided based on the comparison result of two edge cells in the band of score matrix. Specifically, if the value of the rightmost cell is greater than the leftmost cell, this suggests the optimal path is more likely to go rightward \cite{suzuki2017acceleration}. Hence, the current wavefront is moved rightward. Otherwise, the wavefront is moved downward. The adaptive wavefront direction scheme only needs one comparison each iteration but effectively improves the accuracy of long-read alignment according to our test results in Table \ref{table:validation}.

\begin{table}[t]
	\scriptsize
	\centering
	\caption{Comparison of DP alignment algorithms in Fig. \ref{fig:adaptive_band}}
	\label{table:comp_alignment}
	\resizebox{\linewidth}{!}{
	\begin{tabular}{|c|c|c|c|c|}
		\hline
		\multirow{2}{*}{\textbf{Algorithm}} & \multicolumn{2}{c|}{\textbf{Complexity}} & \textbf{Critical} & \multirow{2}{*}{\textbf{Accuracy}}\\
		\cline{2-3}
		& \textbf{Computation} & \textbf{Memory} & \textbf{Path} & \\
		\hline
		\textbf{Full DP} & $\mathcal{O}(mn)$ & $\mathcal{O}(mn)$ & $5\times 32$ bit & High\\
		\hline
		\textbf{Banded Difference-based DP} & $\mathcal{O}(mB)$ & $\mathcal{O}(mB)$ & $8\times 5$ bit & Low\\
		\hline
		\textbf{Adaptive Banded Parallelized DP} & $\mathcal{O}(mB)$ & $\mathcal{O}(mB)$ & $4\times 5$ bit & High\\
		\hline
	\end{tabular}
	}
\end{table}

We conduct an algorithmic analysis for the aforementioned DP algorithms and compare their complexity, data parallelism, and critical path in Table \ref{table:comp_alignment}. The critical path is defined as the longest data path needed to accomplish one iteration of cell updating. Thanks to the alignment matrix parallelism, the proposed adaptive banded parallelized alignment only needs half of the critical path of Eq. (\ref{eq:align_diff}). More importantly, the adaptive wavefront direction compensates for the accuracy loss caused by narrow bandwidth, allowing the proposed algorithm to generate near-optimal results using near-linear complexity.

\begin{figure*}[t]
    \centering
	\includegraphics[width=0.93\textwidth]{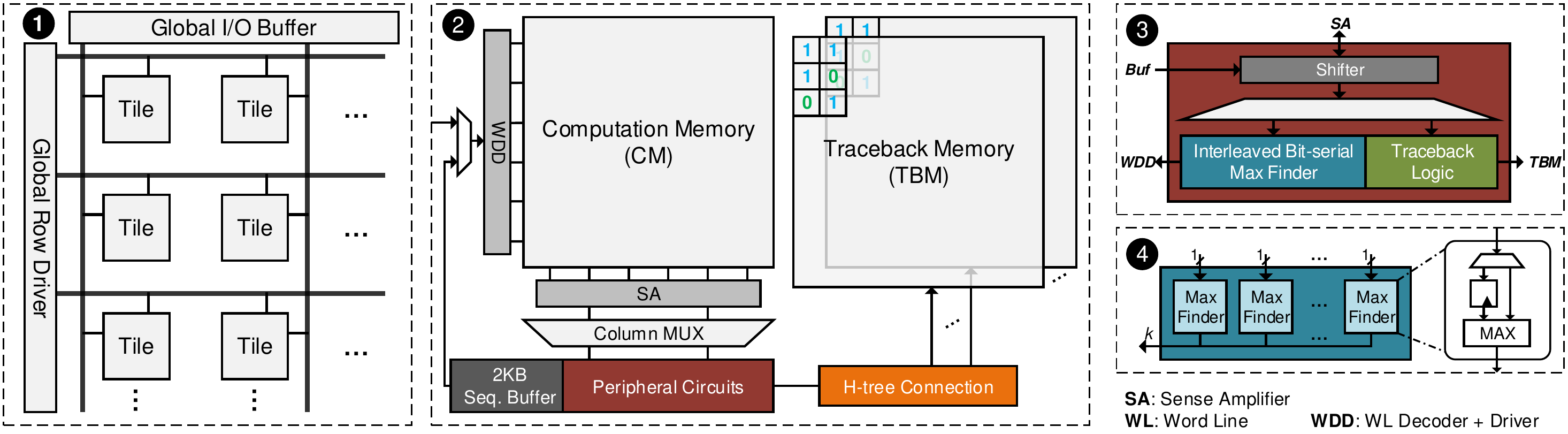}
	\caption{\Design architecture. \encircle{1} ReRAM memory organization of \Design. \encircle{2} Internal architecture of \Design tile. \encircle{3} Peripheral circuits (shifter, interleaved bit-serial max finder, and traceback logic). \encircle{4} Interleaved bit-serial max finder.}
    \vspace{-0.4cm}
	\label{fig:arch_overview}
\end{figure*}

%%%%%%%%%%%%%%%%%%%%%%%%%%%%%%%%%%%%%%%%%%%%%%%%%%%%%%%%%%%%%%%%%%%%%%%%%%%%%%%%%%
\section{In-Memory Architecture of \Design}
We propose the PIM-based ReRAM accelerator, \Design to implement the adaptive banded parallelized DP alignment in Section \ref{sec:diff_align_alg}. \Design~utilizes the in-site PIM-based alignment algorithm and the efficient data flow with four-level parallelism to boost alignment process.

\subsection{Overview} \label{subsec:rapid_overview}
As shown in \encircle{1} of Fig. \ref{fig:arch_overview}, \Design is a ReRAM-based PIM accelerator for genome sequence alignment. The algorithm in Section \ref{subsec:adap_band_para_dp} exhibits various data parallelisms, including wavefront and alignment matrix levels. \Design is organized in a multi-level hierarchy to extend the data parallelism. \Design consists of 64 tiles, each \Design tile independently receiving and transferring genome data through global I/O buffer and global row driver. The read genome sequences are stored in the sequence buffer within each tile. To minimize the data movement, the forward DP cells updating and traceback computation happen locally in each tile. There is no communication between tiles. We conduct design space exploration in Section \ref{subsec:dse} to choose the hardware configurations resulting in the best efficiency.

Fig. \ref{fig:arch_overview}-\encircle{2} shows the internal structure of \Design tile, where one computation memory (CMs) and multiple traceback memories (TBMs) are implemented. One CM is connected to 15 TBMs through the H-tree connection, allowing low-latency and high-bandwidth data transfer between CMs and TBMs. The number of TBM is more than CM because most of the memory is used for storing traceback information. Each CM fetches the reference and query sequences from the 2KB sequence buffer. Then CM calculates $A'$, $\Delta H'$, $\Delta V'$, $\Delta E'$, and $\Delta F'$ matrices in Eq. (\ref{eq:par_align_diff}) using PIM-based XOR and addition operations combined with peripheral circuits. Each CM is able to access TBMs and transfer traceback data through the H-tree routing. Although the ReRAM subarray exhibits high data parallelism, some computations of alignment and traceback can not be efficiently realized in CM. For example, finding the point-wise maximum values of two vectors in~\cite{gupta2019rapid} is complex, requiring both leading one detector and bit-wise logical operations. PIM operations \cite{gupta2018felix} is unable to support low-latency traceback in Eq. (\ref{eq:tb}) as well as the adaptive wavefront direction scheme. In \Design, we connect peripheral circuits to sense amplifier (SA) and offload these operations to the peripheral circuits, consisting of the shifter, interleaved bit-serial max finder, and traceback logic as shown in Fig. \ref{fig:arch_overview}-\encircle{3} and \encircle{4}.

In the peripheral circuits, we identify the max finder accounts for the largest area and has the most complex structure. The design of max finder faces several challenges. First, the additional overhead should be as low as possible to ensure not significantly sacrificing ReRAM memory density. Second, the max finder should match the processing rate of CM while minimally impacting the overall throughput. The max finding scheme in \cite{gupta2019rapid} incurs long latency. We further reduce the latency by offloading the max finding to the interleaved bit-serial max finder in Fig. \ref{fig:arch_overview}-\encircle{4}. The interleaved bit-serial max finder is composed of $k$ bit-serial max finders and the width $k$ equals to the SA's bit width. This is to match the data rate of SA. The classic bit-serial max finder receives 2-bit input in parallel. However, only 1-bit data of multiple data points in the same vector can be read from CM through SA due to CM's bit-serial data organization. Hence, we add a latch and MUX before the input of bit-serial MAX finder to make it support the comparison of bit-serial data.

\subsection{Data Flow with Four-level Data Parallelism} \label{subsec:diff_align_pim}
\begin{figure*}[t]
    \centering
	\includegraphics[width=0.85\linewidth]{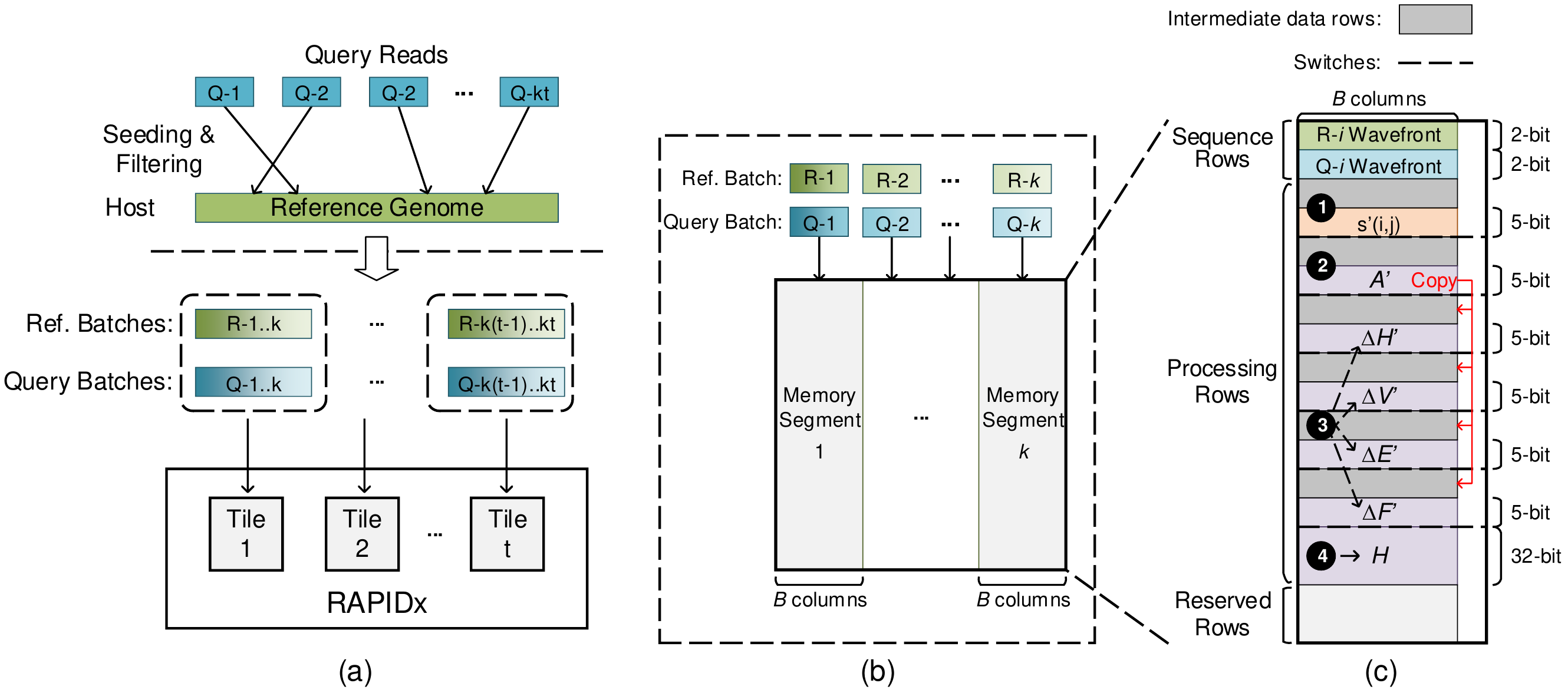}
	\caption{Four-level data parallelism and in-memory alignment in \Design: (a) Tile-level parallelism. (b) Batched alignment in CM using sequence-level parallelism, (c) PIM-based in-situ banded parallelized alignment in each memory segment of CM.}
	\label{fig:data_flow}
    \vspace{-0.2cm}
\end{figure*}

To fully exploit the acceleration opportunities and increase throughput, \Design achieves four-level parallelism, namely tile level, sequence level, wavefront level, and alignment matrix level, as illustrated in Fig. \ref{fig:data_flow}. On the host side, query reads are seeded and filtered in a batched processing manner. Then the resulted $kt$ batches of reference and query pairs are sent to \Design, where $k$ denotes the number of memory segments in Fig. \ref{fig:data_flow} (b) and $t$ denotes the number of tiles. The $kt$ batches of reference and query data are evenly distributed to each tile. The tile-level parallelism enables different \Design tiles to process and align $k$ independent sequences in parallel, allowing the performance of \Design to scale almost linearly with the number of implemented tiles. The CM subarray with size $1024\times 1024$ used in this paper introduces long latency due to the slow PIM operations \cite{gupta2018felix}. The genome sequences in each CM are processed in batch to amortize the long latency of PIM. As illustrated in Fig. \ref{fig:data_flow} (b), each CM processes a reference and a query batch with batch size $k$. The CM is horizontally divided into $k$ memory segments to compute the $k$ pairs of reference and query sequences in parallel. The column width of each memory segment equals the bandwidth $B$ of banded alignment. Hence, there are at most $\lfloor \frac{1024}{B} \rfloor$ memory segments.

\Design achieves wavefront-level and alignment matrix-level parallelism in the memory segments of CM. The wavefront-level parallelism is based on the fact that the cells over anti-diagonal have no data dependency since they only depend on the cells in the previous diagonal. The row-parallel operations of ReRAM subarray compute and update the $B$ wavefront cells over the anti-diagonal in Fig. \ref{fig:adaptive_band} (c) simultaneously. Meanwhile, the relaxed data dependency of parallelized alignment in Eq. (\ref{eq:par_align_diff}) provides the alignment matrix-level parallelism, where $\Delta H_{i,j}'$, $\Delta V_{i,j}'$, $\Delta E_{i,j}'$, and $\Delta F_{i,j}'$ can be computed in parallel.

\subsection{In-memory Alignment}
\subsubsection{Forward DP Updating}
As shown in Fig. \ref{fig:data_flow} (c), the data in ReRAM subarray are organized in the bit-serial manner, where each $b$-bit data lies vertically in $b$ consecutive rows over the bit line. The rows of each memory segment are vertically divided into two regions, including sequence rows and processing rows. The sequence rows are used for storing DNA bases of reference and query. Before starting the wavefront cells updating, the DNA bases related to $B$ wavefront cells are fetched from the sequence buffer and written to the sequence rows. Since each DNA base, $\mathsf{A,G,C,T}$, is encoded with 2-bit data, the sequence rows occupy 4 memory rows. The rest of memory rows work as processing rows and reserved rows, which are responsible for updating wavefront cells of $A_{i,j}', \Delta H_{i,j}', \Delta V_{i,j}', \Delta E_{i,j}', \Delta F_{i,j}'$ in Eq. (\ref{eq:par_align_diff}) using PIM operations \cite{gupta2018felix}. The processing rows are partitioned into five partitions by switches and $A_{i,j}', \Delta H_{i,j}', \Delta V_{i,j}', \Delta E_{i,j}', \Delta F_{i,j}'$ are stored and processed in each partition. Intermediate data rows are inserted into the processing rows to store constant values and intermediate results during computation. The constants used for comparison and subtraction when updating the DP alignment include $2o+2e$ and $o$. These pre-defined values are replicated and pre-stored in the reserved rows. PIM operations can directly access these values whenever needed. Specifically, the forward DP updating is computed in the following orders:  
\begin{enumerate}[leftmargin=*]
	\item First, the 5-bit data $s'(i,j)$ are computed by comparing reference and query wavefront sequences (see \encircle{1} of Fig. \ref{fig:data_flow} (c)). $s'(i,j)$ requires one comparison and addition to generate the match or mismatch score. The comparison between genome bases is done using 2-bit XOR operations.
	\item Second, $A'_{i,j}$ is obtained from the maximum value of $s'(i,j)$, $\Delta E'_{i-1,j}$, and $\Delta F_{i,j-1}$ as shown in \encircle{2} of Fig. \ref{fig:data_flow} (c). Two $\max$ operations are needed in this step.
	\item Third, four copies of $A'_{i,j}$ are written to the intermediate data rows related to $\Delta H_{i,j}', \Delta V_{i,j}', \Delta E_{i,j}'$, and $\Delta F_{i,j}'$ as \encircle{3} of Fig. \ref{fig:data_flow} (c).
	\item Third, $\Delta H_{i,j}'$ and $\Delta V_{i,j}'$ are updated in parallel using copied $A_{i,j}$ and previous $\Delta V_{i-1,j}', \Delta H_{i,j-1}'$. Meanwhile, $\Delta E_{i,j}'$ and $\Delta F_{i,j}'$ are updated in parallel based on copied $A_{i,j}'$ and $\Delta E_{i-1,j}'$, $\Delta F_{i,j-1}'$, $\Delta H_{i,j-1}'$, $\Delta V_{i-1,j}'$ of the previous iteration. This step needs four subtractions, two additions, and two $\max$ operations.
	\item Finally, the alignment score matrix $H_{i,j}$ need to be retrieved using the function $H_{i,j} = H_{i-1,j} + \Delta H_{i,j} = \Delta H_{i,j}'-(o+e) + H_{i-1,j}$, which requires one 5-bit subtraction and one 32-bit addition.
\end{enumerate}

\begin{figure}[ht]
    \centering
	\includegraphics[width=0.8\linewidth]{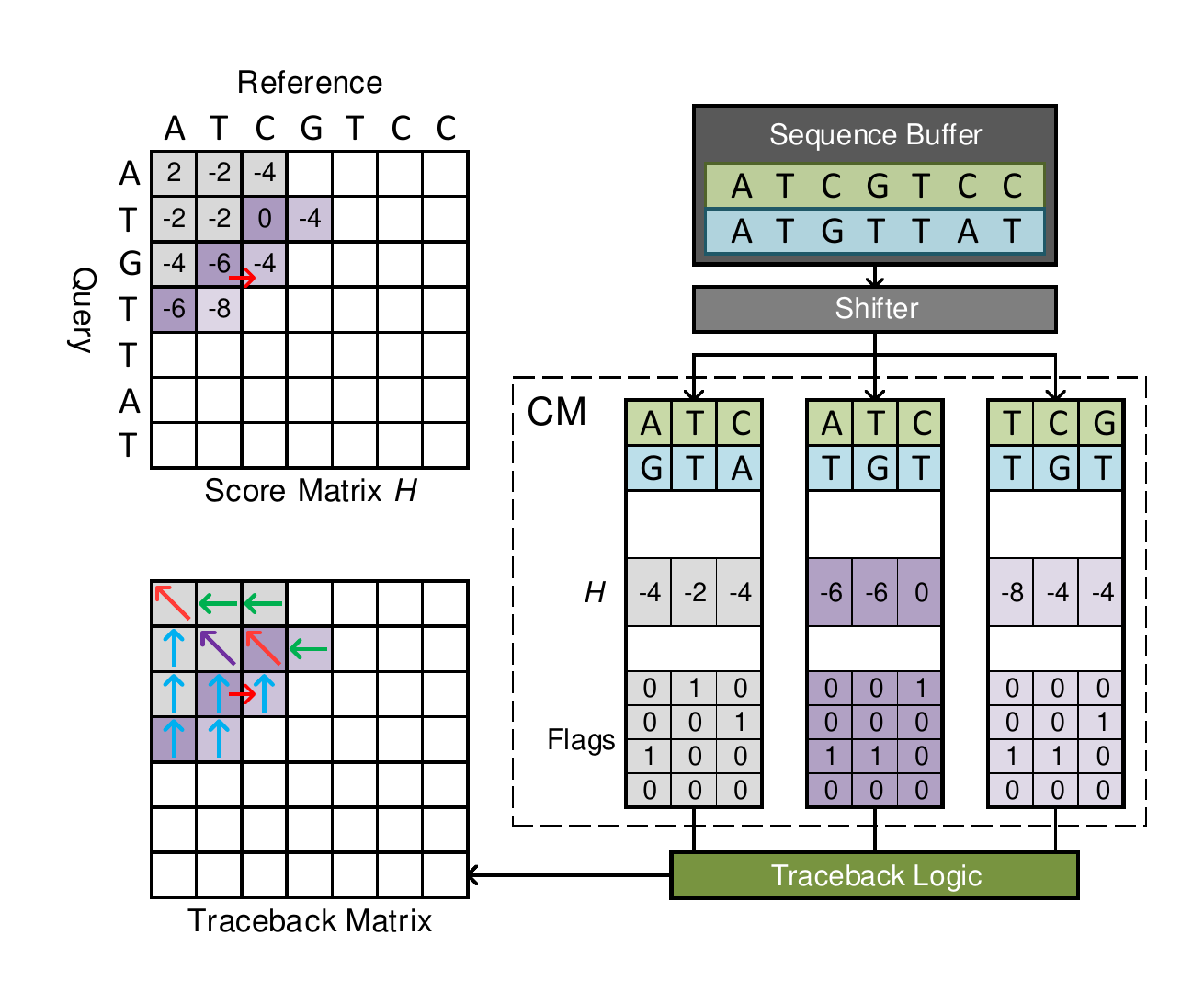}
	\caption{Illustration of adaptive wavefront direction and traceback process using peripheral circuits.}
	\label{fig:adaptive_wavefront_tb}
\end{figure}

% \begin{figure}[ht]
% 	\centering
% 	\includegraphics[width=0.75\linewidth]{Figures/adaptive_wavefront_tb.pdf}
% 	\caption{Illustration of adaptive wavefront direction and traceback using peripheral circuits.}
% 	\label{fig:adaptive_wavefront_tb}
% \end{figure}

\subsubsection{Adaptive Wavefront Direction}
After wavefront cells are computed, the band will move either downwards or rightwards by one cell. Fig. \ref{fig:adaptive_wavefront_tb} illustrates how the wavefront with bandwidth $B=3$ is moved using peripheral circuits, where the wavefront direction is controlled by the shifter and sequence buffer. The max finder first compares the leftmost and rightmost cells in score matrix $H$, determining the next direction for wavefront. Then, the shifter receives the direction signal and reads the corresponding genome sequence from the sequence buffer. If the wavefront is moving rightwards, the shifter fetches reference data. Otherwise, it fetches query data. After shifting to the position of wavefront cells, the new genome sequence is written to the sequence rows within CM. In this way, the majority of computation data stay stationary in CM using in-situ PIM-based alignment, reducing the data movement overhead.

\subsubsection{Traceback Process}
Each iteration of DP alignment is followed by updating traceback matrix. Eq. (\ref{eq:align_affine}) can easily compute the traceback matrix through comparing the corresponding values of three alignment matrices $I$, $D$, and $H$. However, the difference-based DP alignment in Eq. (\ref{eq:align_diff}) and Eq. (\ref{eq:par_align_diff}) only store the difference values and do not explicitly give the score matrix $H$. Therefore, we modify the formula of generating traceback information of the original DP to calculate the traceback matrix $TB$ as the following equation:
\begin{align}\label{eq:tb}
    \footnotesize
    	TB_{i-1,j-1} = 
    	\begin{cases}
    		00, \text{if\ } s'_{i,j} == (A+o+e)~\text{or\ } (-B+o+e)\\
    		01, \text{if\ } \Delta H_{i,j}' == \Delta E_{i-1,j}' - \Delta V_{i-1,j}' \\ 
    		10, \text{if\ } \Delta H_{i,j}' == \Delta F_{i,j-1}' - \Delta H_{i,j-1}'\\ 
    		11, \text{if\ others} 
    	\end{cases}
\end{align}
where two subtractions and four comparisons are needed. $00$, $01$, and $10$ denote the cases of match or mismatch, deletion, and insertion, respectively. 

As shown in Fig. \ref{fig:adaptive_wavefront_tb}, to efficiently implement Eq. (\ref{eq:tb}) in memory, the traceback logic in \encircle{4} of Fig. \ref{fig:arch_overview} reads out the 4-bit flags that indicate the traceback information from CM in a bit-serial order. Then the traceback logic converts the 4-bit flags into 2-bit traceback data and stores them into TBM. Since there will be only one ``1" in the 4-bit flags. The conversion from 4-bit flags to 2-bit data is accomplished by implementing one hot encoders within the traceback logic.

\subsection{Reconfigurable Design with Dynamic Precision}\label{subsec:dyn_precision}
The sequence alignment and edit distance calculation follow the same data flow of forward cell updating. The difference between alignment and edit distance calculation is the used scoring function. The scoring function of edit distance computation normally requires lower data bit width than alignment workloads. \Design is reconfigurable to support these two workloads by adopting two types of PIM precisions. Moreover, we leverage the precision difference to further improve the performance of edit distance calculation. 

\subsubsection{Alignment Computation}	
For different alignment tools and target genomes to be aligned, various scoring functions with affine gap penalties may be applied. For example, BWA-MEM \cite{bwamem} uses a matching score $A=1$, mismatch penalty $B=4$, gap open penalty $o=6$, and gap extension penalty $e=1$. The other popular alignment tool, Minimap2 \cite{minimap2}, uses a default scoring function with $A=2, B=4, o=4, e=2$. According to Section \ref{sec:diff_align_alg}, the minimum data width should satisfy $\lceil \log_{2}(M+2o+2e+1)\rceil$. For most scoring functions with affine gap penalties, a 5-bit PIM precision is able to realize accurate alignment without overflow.

\subsubsection{Edit Distance Calculation}
Edit distance (or Levenshtein distance) is a metric to measure the minimum number of deletion, insertion, and substitution required to transform one string to the other one.	Edit distance calculation can be regarded as a simplified version of sequence alignment, where the matching score is 0 while mismatch/gap opening/gap extension penalties are all 1. $\lceil \log_{2}(M+2o+2e+1)\rceil = 3$-bit data width provides sufficient precision for edit distance calculation. Therefore, \Design decreases the arithmetic precision from 5-bit to 3-bit when computing edit distance. This is beneficial to further improve throughput and reduce energy dissipation.

\Design realizes the switching between the mentioned two types of PIM precisions through issuing different sets of commands to CMs. The commands for 3-bit and 5-bit precisions differ in they activate and access different ReRAM rows in CM to realize different computing precisions. So the overhead of PIM precision switching is negligible.

%%%%%%%%%%%%%%%%%%%%%%%%%%%%%%%%%%%%%%%%%%%%%%%%%%%%%%%%%%%%%%%%%%%%%%%%%%%%%%%%%%%%%%%%%%
%%%%%%%%%%%%%%%%%%%%%%%%%%%%%%%%%%%%%%%%%%%%%%%%%%%%%%%%%%%%%%%%%%%%%%%%%%%%%%%%%%%%%%%%%%

\section{Evaluation}
\subsection{Experimental Setup}
\noindent
\textbf{Methodology:}
We use VTEAM \cite{kvatinsky2015vteam} with $R_{OFF}=300k$ and $R_{ON}=10k$ to model ReRAM cell. The other parameters are same with \cite{talati2016logic} that align with the practical ReRAM device \cite{yang2013memristive}. The energy consumption and latency of PIM operations in \Design are measured based on 10,000 Monte Carlo simulations in SPICE. The operation voltage of PIM is $V_0$=1V, and the worst-case switching latency is 2ns. The hardware parameters of ReRAM subarray are obtained from NVSim \cite{dong2012nvsim}. Its peripheral circuits, including shifter, interleaved bit-serial max finder, and traceback logic, are implemented using \textit{Verilog} and synthesized by \textit{Synopsys Design Compiler} on 45nm process node \cite{stine2007freepdk}. The area and energy consumption of sequence buffer are estimated using CACTI \cite{cacti}. \Design's frequency is set to 500MHz, matching the switching time of ReRAM device. We also develop a in-house simulator to estimate the DNA alignment performance and energy consumption.

\noindent
\textbf{\Design Configurations:}
Total 64 tiles are implemented in \Design and each \Design tile has 2MB size, containing one CM and 15 TBMs. Each ReRAM subarray consists of $1024\times 1024$ cells and the width of column MUX output is 128-bit. The parameter selection is discussed in Section \ref{subsec:dse}. The arithmetic precision is set to 5-bit for sequence alignment and 3-bit for edit distance calculation, which avoids overflow and maximizes the performance.

\begin{table}[ht]
	\scriptsize	
	\caption{Error rates of generated datasets}
	\centering
	\begin{tabular}{|c|c|c|c|c|}
		\hline
		\textbf{Type} & \textbf{Substitution} & \textbf{Insertion} & \textbf{Deletion} & \textbf{Total}\\
		\hline
		\textbf{PacBio} & 1.5\% & 9.0\% & 4.5\% & 15\%\\
		\hline
		\textbf{ONT\_2D} & 16.5\% & 5.0\% & 8.5\% & 30\%\\
		\hline
		\textbf{Illumina} & 3\% & 1\% & 1\% & 5\%\\
		\hline
	\end{tabular}\label{table:error_profile}
\end{table}

\noindent
\textbf{Datasets:}
We test \Design's performance on both short and long reads. The sequence length of short reads ranges from 100bp to 500bp while the long reads vary from 2kbp to 10kbp. We use the homologous chromosomes, GRCh38 \cite{GRCh38}, from the National Center for Biotechnology Information (NCBI). The chromosomes, including 1 to 22, X, and Y, are used and the unmapped contigs are removed. These chromosomes contain 3 billion bp in total. The available memory space in \Design is not able to store the entire genome. We assume \Design fetches the query and reference sequences from the host memory for alignment.

\begin{table}[ht]
	\scriptsize	
	\caption{Hardware specifications of CPU and GPU baselines}
	\centering
	\resizebox{\linewidth}{!}{
	\begin{tabular}{|c|c|c|c|}
		\hline
		\multirow{2}{*}{\textbf{CPU}} & Intel Xeon E5-2680 & \multirow{2}{*}{\textbf{GPU}} & \multirow{2}{*}{\textbf{Geforce GTX 1080 Ti}} \\
		\cline{2-2}
		& 12 cores / 24 threads / 2.5GHz  & & \\
		\hline
		\textbf{Cache} & L1/L2/L3: 32KB/256KB/30MB & \textbf{Frequency} & 1582 MHz \\
		\hline
		\textbf{Memory} & 256GB / DDR4-2133MHz & \textbf{Memory} & 11GB GDDR5X\\
		\hline
		\textbf{TDP} & 120W & \textbf{TDP} & 250 W \\
		\hline
	\end{tabular}
	}
	\label{table:cpu_gpu_baselines}
\end{table}

As Table \ref{table:error_profile}, we generate the long-read data (PacBio and ONT datasets) using the sequence read simulator PBSIM \cite{ono2013pbsim}. PacBio and ONT have 15\% and 30\% error rate, respectively. PBSIM's default error profile and continuous long read (CLR) mode are used. The short-read Illumina datasets are produced by Mason \cite{mason} with 5\% error rate. Both \Design and other baselines are tested using at least 100,000 reads for each length.

\begin{table}[ht]
	\scriptsize	
	\caption{Specifications of ASIC baselines}
	\centering
	\resizebox{\linewidth}{!}{
	\begin{tabular}{|c|c|c|}
		\hline
		\textbf{Design} & \textbf{ABSW \cite{liao2018adaptively}} & \textbf{GenASM \cite{liao2018adaptively}}\\
		\hline
		\multirow{2}{*}{\textbf{Specifications}} & 40nm with 480MHz frequency & 28nm with 1GHz frequency\\
        \cline{2-3}
		 & Area: 5.51mm$^{2}$, Power: 1.2W & Area: 10.69mm$^{2}$, Power: 3.2W\\
		\hline
	\end{tabular}
	}
	\label{table:asic_baselines}
\end{table}

\noindent
\textbf{Baselines:}
We compare the alignment performance of	\Design with state-of-the-art CPU, GPU, PIM, and ASIC accelerators. The CPU baselines include two libraries developed using C++, Minimap2 \cite{minimap2} and Edlib \cite{vsovsic2017edlib}. Minimap2 utilizes banded DP algorithms with affine gap penalties and adopts SIMD and multithreading to maximize the performance. Edlib is a C++ program that makes use of edit distance and Myers's bit-vector algorithm \cite{myers1999fast} to parallelize the alignment and distance computation. The GPU baseline, GASAL2 \cite{gasal2}, is optimized for GPU and delivers high throughput on various alignment workloads. We compile and run the programs on a server with hardware specifications in Table \ref{table:cpu_gpu_baselines}. The other parameters are the same as the original papers \cite{minimap2,vsovsic2017edlib,gasal2} without explicit specifications. 
We compare \Design with four PIM designs, including RAPID \cite{gupta2019rapid}, AlignS \cite{angizi2019aligns}, AligneR \cite{zokaee2018aligner}, and PIM-Aligner~\cite{angizi2020pim}. We also compare \Design with two ASIC accelerators, ABSW \cite{liao2018adaptively} and GenASM \cite{genasm}. Their hardware configurations are given in Table \ref{table:asic_baselines}.

% ABSW adopts the adaptive banded DP alignment algorithm with affine gap penalties based on 12-bit integers. Instead of using the DP-based alignment algorithms, GenASM exploits a modified approximate string matching algorithm to increase the parallelism and reduce memory footprint. Table \ref{table:asic_baselines} summarizes the area, frequency, and power consumption of ABSW and GenASM.

\subsection{Algorithm Validation}\label{subsec:alg_valid}
The bandwidth of adaptive banded DP alignment is key to the alignment accuracy and efficiency. The base bandwidth $w$ in the bandwidth calculation function $B=\min(w + 0.01\times L, 100)$ determines the resulted bandwidth for sequence length $L$. Large $w$ guarantees high alignment accuracy but increases the required computation and memory complexity.

\begin{table}[ht]
	\scriptsize
	\caption{Alignment accuracy of banded DP algorithms}
	\centering
	\resizebox{\linewidth}{!}{
	\begin{tabular}{|c|c|c|c|c|c|c|}
		\hline
		\multirow{2}{*}{\textbf{Read Type}} & \textbf{Adaptive} & \multicolumn{5}{c|}{\textbf{Base bandwidth $w$}}\\
		\cline{3-7}
		& \textbf{Wavefront} & 10 & 20 & 30 & 40 & 50\\
		\hline
		\textbf{Short Read} & No & 100.0\% & 100.0\% & 100.0\% & 100.0\% & 100.0\%\\
		\cline{2-7}
		\textbf{(Illumina)} & Yes & 100.0\% & 100.0\% & 100.0\% & 100.0\% & 100.0\%\\
		\hline
		\textbf{Long Read} & No & 6.51\% & 39.69\% & 31.33\% & 61.44\% & 71.13\%\\
		\cline{2-7}
		\textbf{(ONT\_2D)} & Yes & 99.23\% & 99.64\% & 99.85\% & 99.85\% & 99.95\%\\
		\hline
	\end{tabular}
	}
	\label{table:validation}
\end{table}

We perform Monte Carlo simulations to validate the accuracy of adaptive banded parallelized DP alignment using different parameters. The alignment results of original DP with affine gap penalty in Eq (\ref{eq:align_affine}) are regarded as the ground truth. Both of the tested algorithm adopt the identical scoring function $A=2, B=4, o=4, e=2$ with Minimap2 \cite{minimap2}. We randomly sample 1,000,000 short and long sequence reads from the read simulator. Illumina and ONT\_2D in Table \ref{table:error_profile} are adopted as the reading scheme for short reads and long reads, respectively. 

Table \ref{table:validation} gives the alignment accuracy, where the base bandwidth $w$ is ranging from 10 to 50 and the bandwidth is calculated by $B=\min(w + 0.01\times L, 100)$. We also add another dimension that enables or disables the adaptive wavefront direction. The results show that the accuracy for short read is all $100\%$ even without adaptive wavefront direction. This is because Illumina only incurs 5\% error. For long reads, the algorithm without adaptive wavefront direction yields unsatisfactory accuracy. Increasing $w$ to 50 only yields 71.13\% accuracy. This is because ONT\_2D has lower reading quality, making the optimal alignment path more likely to be away from the diagonal. The fixed wavefront direction is unable to track and cover the optimal path. After enabling adaptive wavefront direction, a base bandwidth $w$ of 10 achieves 99.23\% accuracy. It is observed that the optimal $w$ varies for reading schemes and sequence lengths. To balance alignment efficiency and accuracy, we choose $w=10$ for short reads and $w=30$ for long reads, which incurs 0.15\% accuracy degradation.

\subsection{Design Space Exploration}\label{subsec:dse}

    \begin{figure}[t]
    	\centering
    	\includegraphics[width=0.85\linewidth]{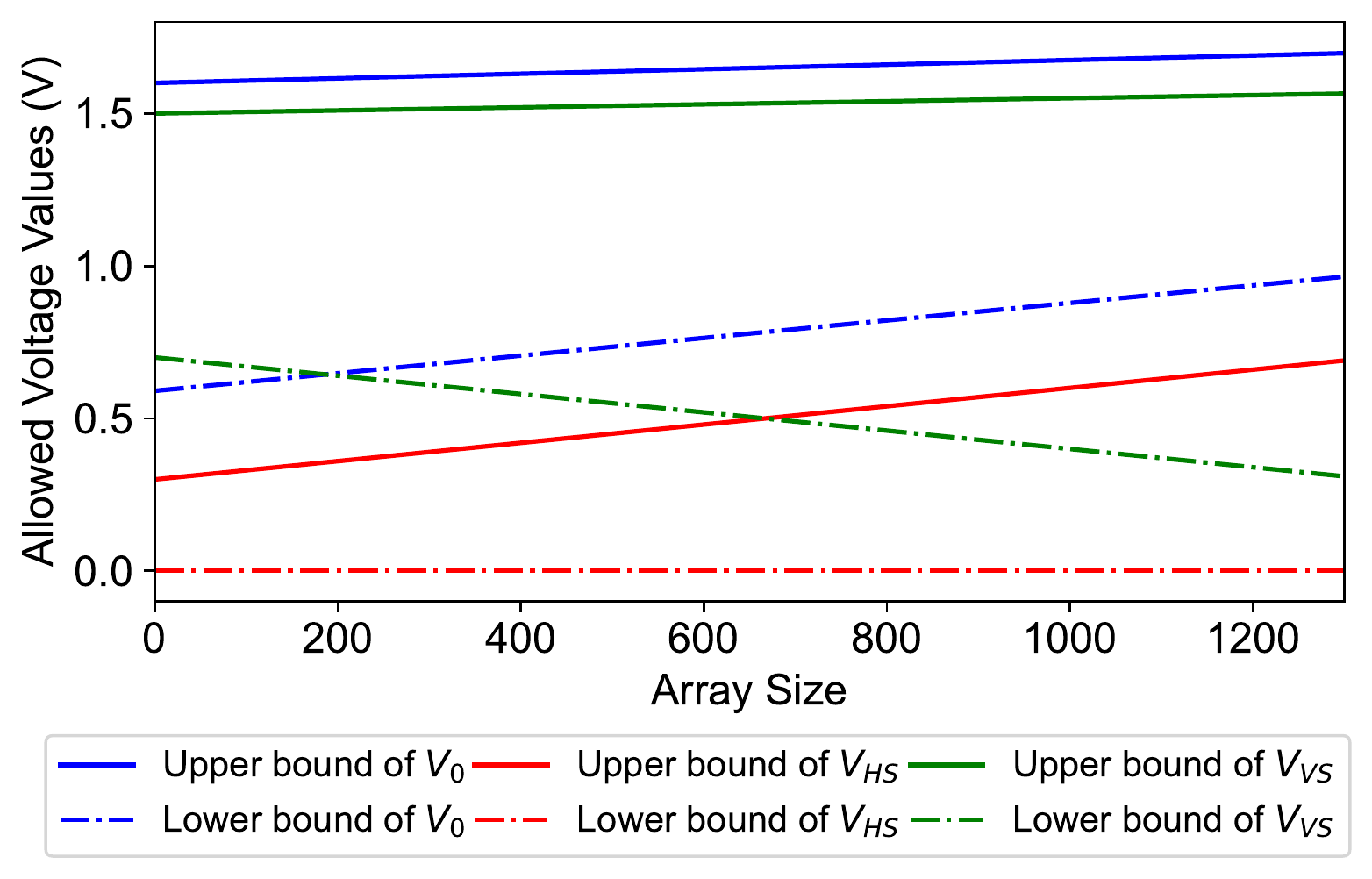}
    	\caption{The lower bound and upper bound of voltages $V_0$, $V_{HS}$, $V_{VS}$ under different ReRAM array sizes.}
    	\label{fig:wire_res_size}
    \end{figure}

    \subsubsection{ReRAM Subarray Size} 
    The ReRAM subarray size determines the memory density. The parasitic wire resistance is a major factor limiting the ReRAM size~\cite{talati2016logic}. To study the impact of non-ideal wire resistance, we use the same model in \cite{talati2016logic} and assume the unit wire resistance between row or column is $R_w=10\Omega$. 
    The upper bound and lower bound of three critical voltages (operation voltage $V_0$, isolation voltages $V_{HS}$ and $V_{VS}$) under different ReRAM array size are depicted in Figure~\ref{fig:wire_res_size}. 
    It shows the used $V_0=1.0$V falls in the allowed value range when array size is $1024\times 1024$. 
    The effective ranges for voltages $V_{HS}$ and $V_{VS}$ show we can set the isolation voltages to $V_{HS}=0.2V, V_{VS}=1.0V$ to satisfy the constraints for size $1024\times 1024$. 
	Given these results, the wire resistance does not affect the correct functionality of \Design under ReRAM array size $1024\times 1024$. This is because: 1. RAPIDx uses 2-input PIM operation to perform alignment, reducing the effects of wire resistance. 2. The 10k$\Omega$ $R_{ON}$ is $10\times$ larger than~\cite{talati2016logic}, making \Design receive less impact from the wire resistance. 
    Meanwhile, the chip-verified ReRAM~\cite{xue202116} with 1024 dimension also demonstrates that the ReRAM subarray in \Design is practical to manufacture.

    \begin{figure}[t]
    	\centering
    	\includegraphics[width=0.9\linewidth]{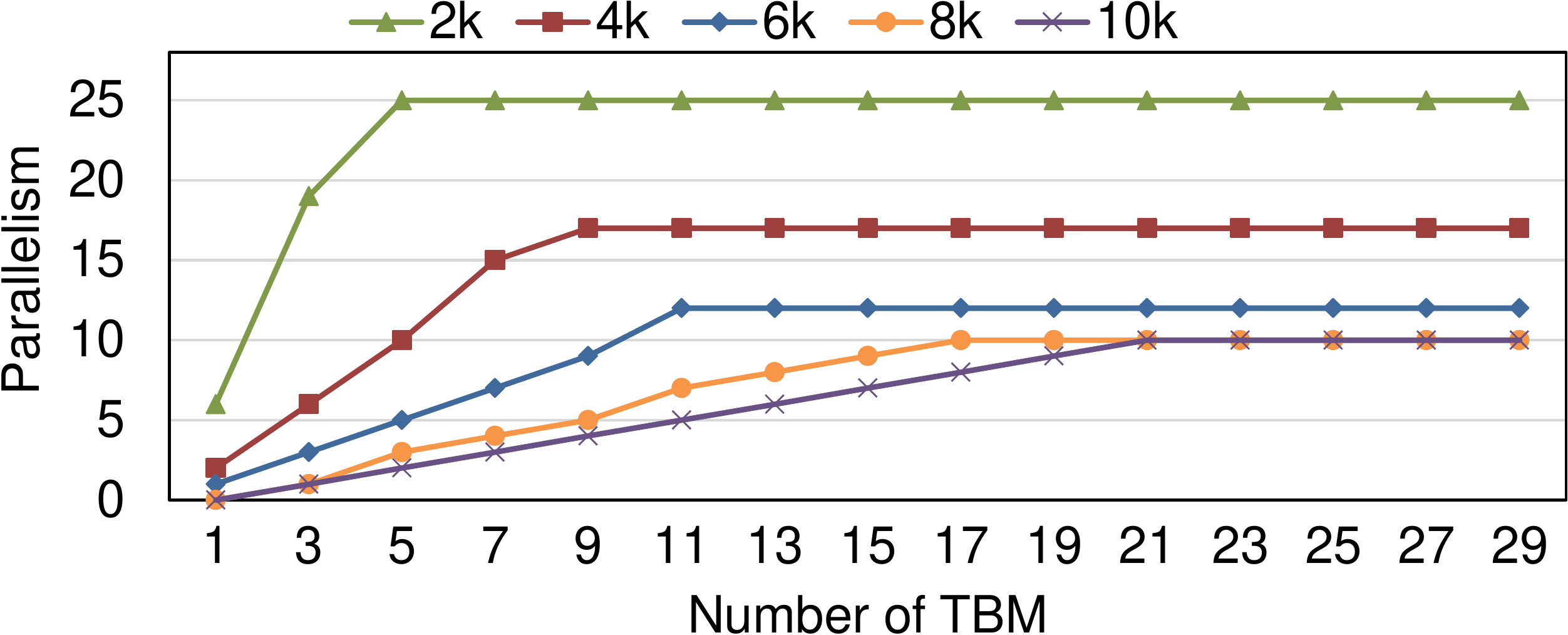}
    	\caption{Relationship between maximum sequence-level parallelism and number of TBMs on long reads.}
    	\label{fig:parallelism_tbm}
    \end{figure}

\subsubsection{Number of TBMs in Each Tile}
The memory complexity of alignment is dominated by traceback data storage because the traceback data for a batch of sequences need to be stored until all DP alignment steps are finished. Therefore, each CM can access the memory space of $t$ TBMs. The number of TBMs in each tile determines the supported maximum sequence length of \Design. Each TBM is a $1024\times 1024$ ReRAM subarray, thus each TBM can store $\frac{1024^2}{2}$ points of traceback data, where 2 denotes the 2-bit traceback information. Considering the sequence alignment or edit distance calculation has a bandwidth $B$ and sequence length $m$, the number of TBMs $t$ in each tile, satisfies $m\le \frac{1024^2}{2B} t$. However, the memory requirement increases linearly by $k\times$ when each CM processes $k$ sequences in parallel. In this case, the maximum sequence level parallelism (or the memory segment) becomes $k \le \lfloor \frac{1024^2}{2m\cdot B} t \rfloor$. On the other hand, $k$ will not exceed the maximum segment number in each ReRAM subarray $k\le \lfloor \frac{1024}{B} \rfloor$. Therefore, the relationship between number of TBMs $t$, sequence-level parallelism $k$, and sequence length $m$ is given by $k\le \min(\lfloor \frac{1024}{B} \rfloor, \lfloor \frac{1024^2}{2m\cdot B} t \rfloor)$.

The sequence-level parallelism under various sequence lengths and TBM numbers is given in Fig. \ref{fig:parallelism_tbm}. Shorter sequences require less TBMs to achieve the maximum parallelism. The $k_{\text{max}}$ of sequences longer than 8kbp is limited by $\lfloor \frac{1024}{B} \rfloor$. As the maximum value of $B$ is 100, $\lfloor \frac{1024}{B} \rfloor \le 10$ for sequences over 8kbp. In this case, the number of TBMs $t$, making $\lfloor \frac{1024^2}{2m\cdot B} t \rfloor > 10$, can not further improve the performance. We implement $t=15$ TBMs to ensure sufficient sequence-level parallelism for 10kbp while balancing area overhead. Thus, each \Design tile consists of 16 ReRAM subarrays.

\begin{figure}[t]
	\centering
	\begin{subfigure}{.4\textwidth}
		\centering
		\includegraphics[width=0.9\linewidth]{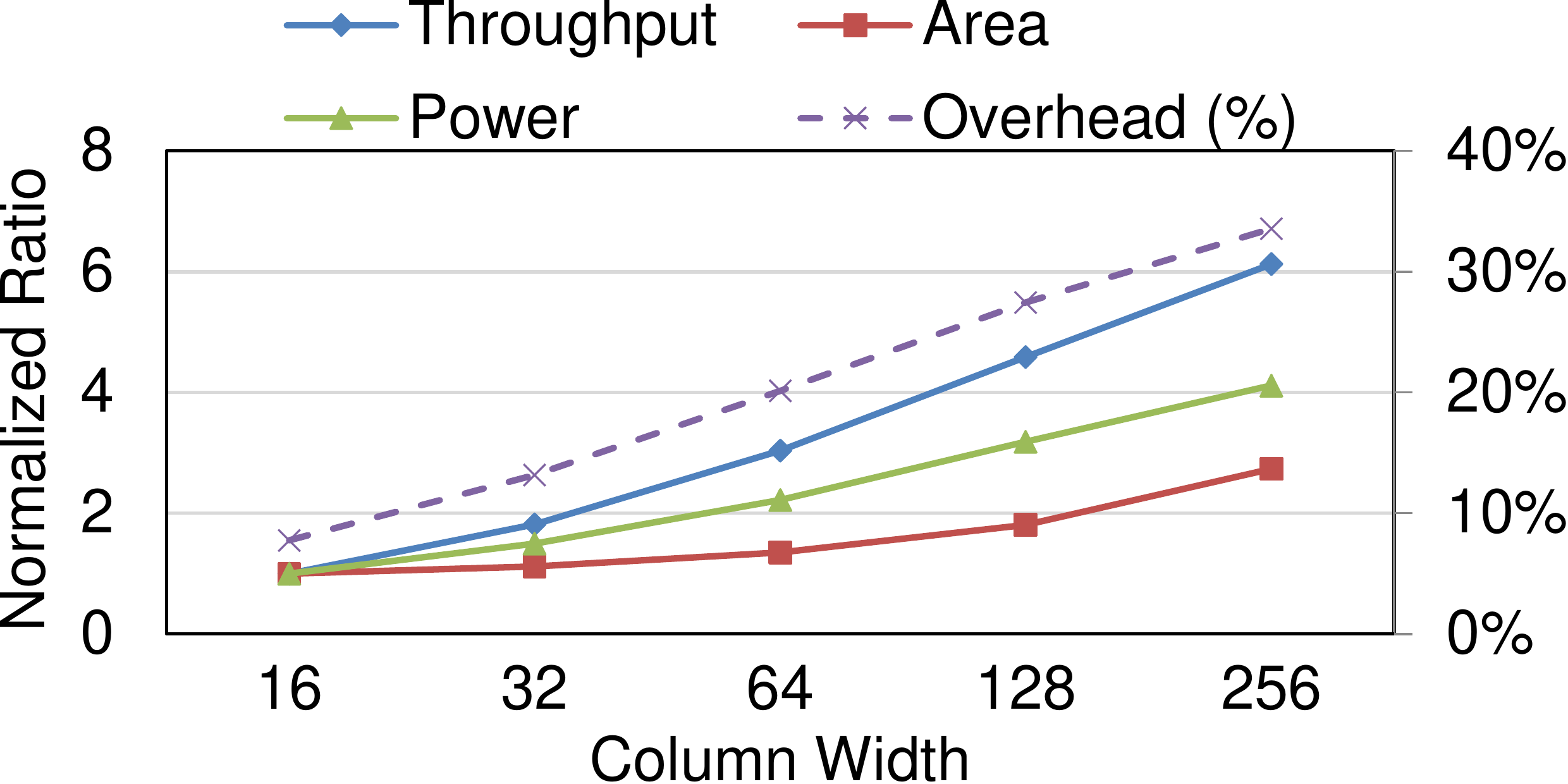}
		\caption{Throughput, area, power, and overhead.}
		\label{fig:dse_1}
	\end{subfigure}%
	\hfil
	\begin{subfigure}{.4\textwidth}
		\centering
		\includegraphics[width=0.9\linewidth]{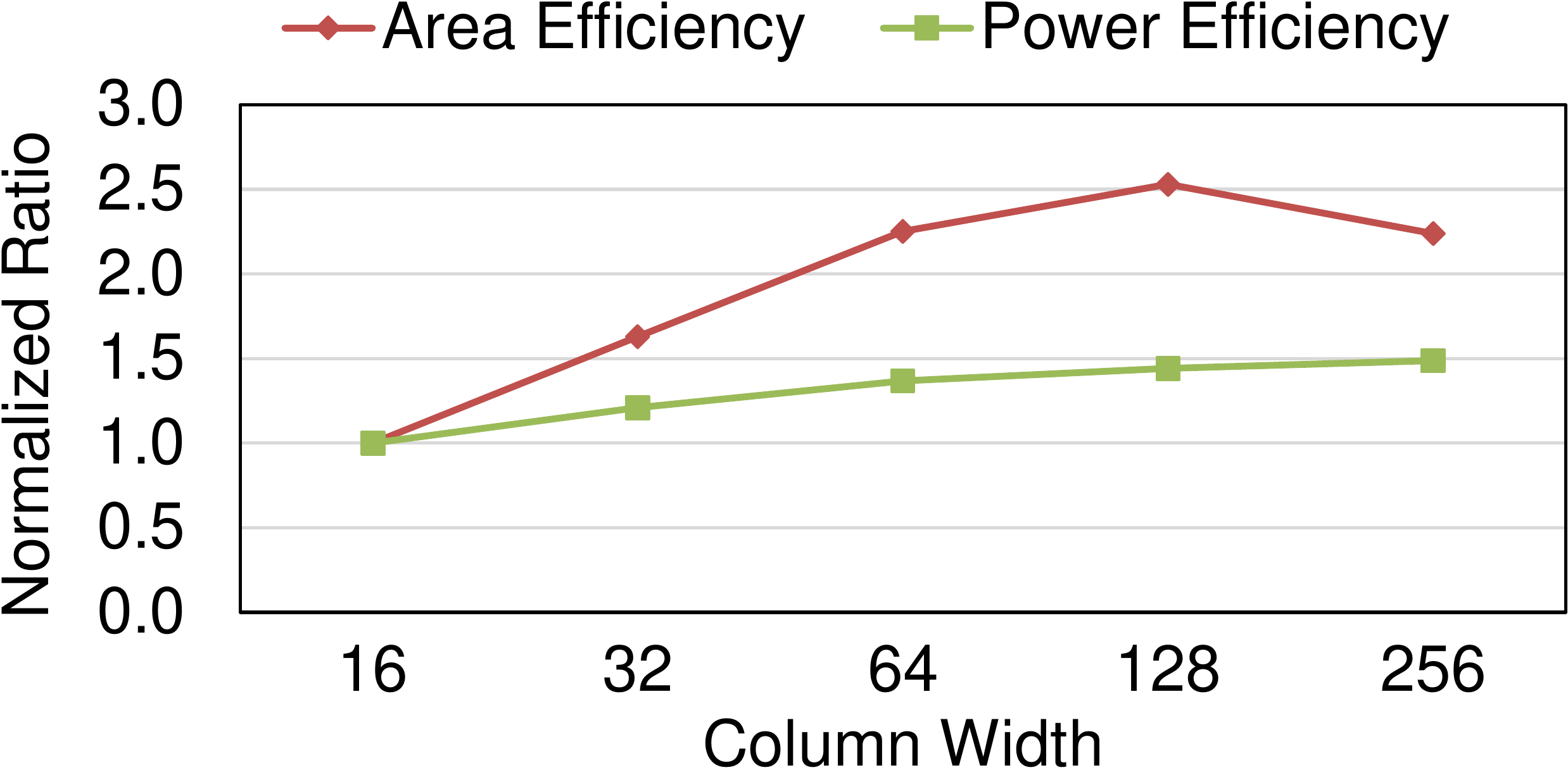}
		\caption{Area efficiency and power efficiency.}
		\label{fig:dse_2}
	\end{subfigure}
	\caption{Performance comparison for different column widths of peripheral circuits.}
	\label{fig:dse}
\end{figure}

\subsubsection{Column Width of Peripheral Circuits}
The peripheral circuits of CM are connected to the column MUX of SA and have the same width as column MUX. The column width of peripheral circuits is a design parameter affecting the overall throughput, power, and area. Fig. \ref{fig:dse} shows the comparison of performance for different widths (from 16 to 256) of peripheral circuits. As shown in Fig. \ref{fig:dse_1}, wider column width leads to higher throughput and the increasing trend of throughput is slightly more significant than area and power when the width is between 16 and 128. The overhead here denotes the percentage of peripheral circuits area to single ReRAM subarray. We depict the area efficiency and power efficiency in Fig. \ref{fig:dse_2} to understand the relationship between efficiency and column width. Area efficiency and power efficiency peak at width 128 and 256, respectively. However, wider width introduces larger area overhead to CM. We choose the column width of 128 to achieve good tradeoff between efficiency and overhead.

\subsection{Area and Power Results}
The area and power breakdown of \Design is summarized in Table \ref{tab:impl_results}. The bit-serial max finder takes up 62.3\% area and 61.6\% power of the peripheral circuits, respectively. About 16\% area of CM is consumed by peripheral circuits. Each \Design tile is composed of 1 CM and 15 TBMs, consuming 0.0.637mm$^2$ area and 0.16W power. We measure the power dissipation of \Design under sequence alignments for long sequence lengths (2kbp to 10kbp) with enabling the traceback procedure. As a result, the area and power of \Design with 64 tiles in total are 40.8mm$^2$ and 10.3W, respectively.

\begin{table}[t]
    \centering
	\scriptsize	
	\caption{Area and power breakdown of \Design}
		\begin{tabular}{|c|r|r|}
			\hline
			\multirow{2}{*}{\textbf{Peripheral Circuits}} & \textbf{Area} & \textbf{Power}\\
			 & \textbf{(um$^2$)} & \textbf{(mW)}\\
			\hline
			Shifter & 542.6 & 0.03 \\
			Max Finder & 4, 520.8 & 2.05 \\
			Traceback Logic & 1, 872,4 & 1.21\\
			Others & 325.2 & 0.03\\
			\hline
			\textbf{Total} & 7, 260.9 & 3.32 \\
			\hline
			\textbf{Seq. Buffer} & 8, 492.6 & 1.5 \\
			\hline
			\hline
			\textbf{ReRAM Subarray} & 38, 395.0 & 9.76\\
			\hline
		\end{tabular}
	    \vfill\vspace{0.1cm}
		\begin{tabular}{|c|r|r|}
			\hline
			\textbf{\Design} & \textbf{Area} & \textbf{Power}\\
			\hline
			\textbf{Per tile} &  637,334.4um$^2$ & 0.16W\\
			\hline
			\textbf{Total} & 40.8mm$^2$ & 10.3W\\
			\hline
		\end{tabular}
	\label{tab:impl_results}
\end{table}

\subsection{Performance Evaluation}\label{subsec:perf_eval}
We measure the performance of \Design on various sequence lengths and compare with state-of-the-art acceleration solutions for genome sequence analysis. The sequences are divided into short reads ($<$1kbp) and long reads ($>$1kbp). Two types of workloads are considered, including sequence alignment in Section \ref{subsec:perf_comp_rapid} \ref{subsec:perf_comp_short} \ref{subsec:perf_comp_long} and edit distance calculation in Section \ref{subsec:perf_comp_edit}. \Design uses 5-bit integer for alignment and 3-bit integer for edit distance calculation.

\begin{figure}[t]
	\centering
	\begin{subfigure}{\linewidth}
		\centering
		\includegraphics[width=0.7\linewidth]{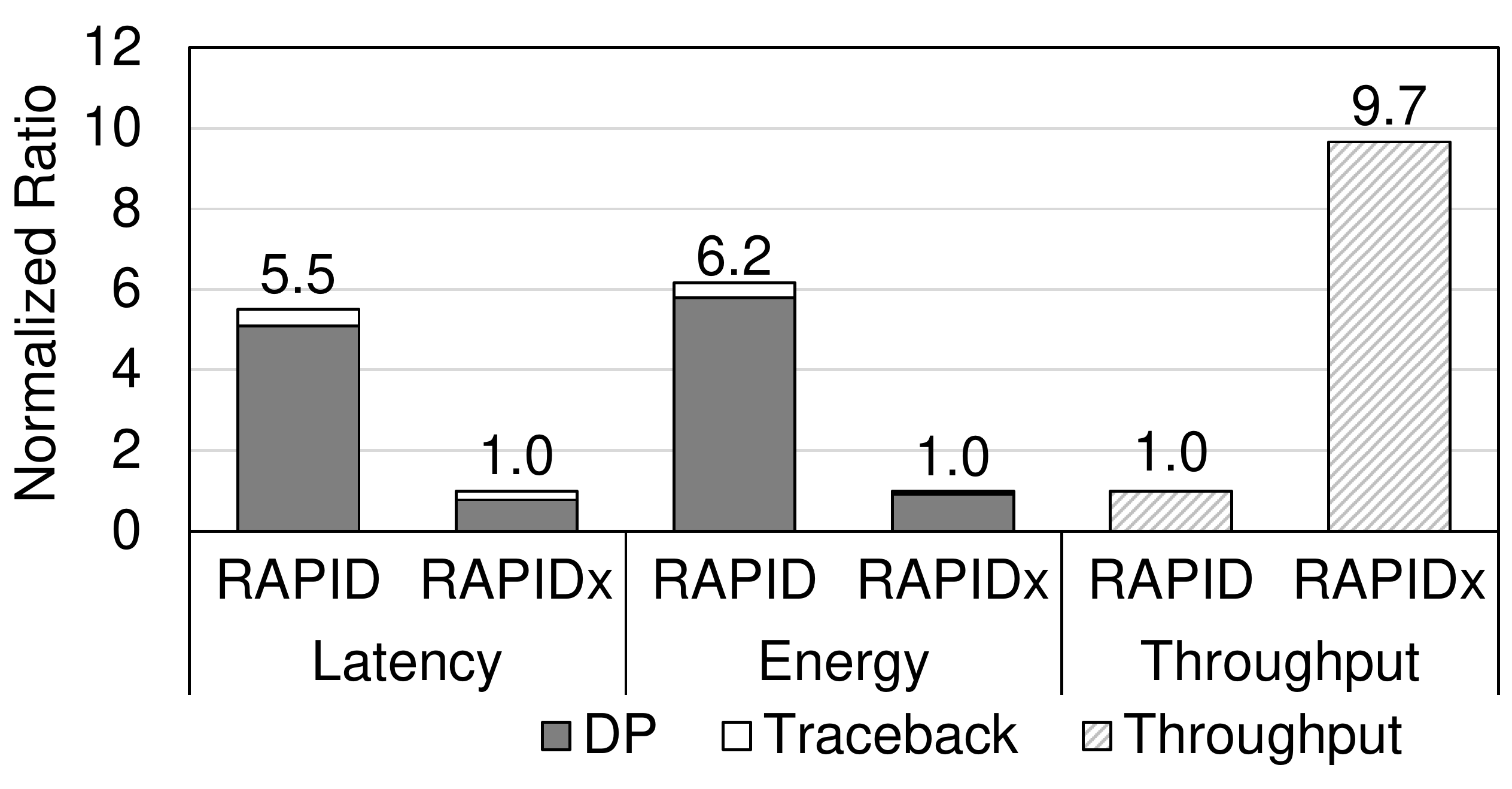}
		\caption{Latency, energy, and throughput comparison of RAPID \cite{gupta2019rapid} and \Design.}
	\end{subfigure}
	\begin{subfigure}{\linewidth}
		\centering
		\includegraphics[width=0.75\linewidth]{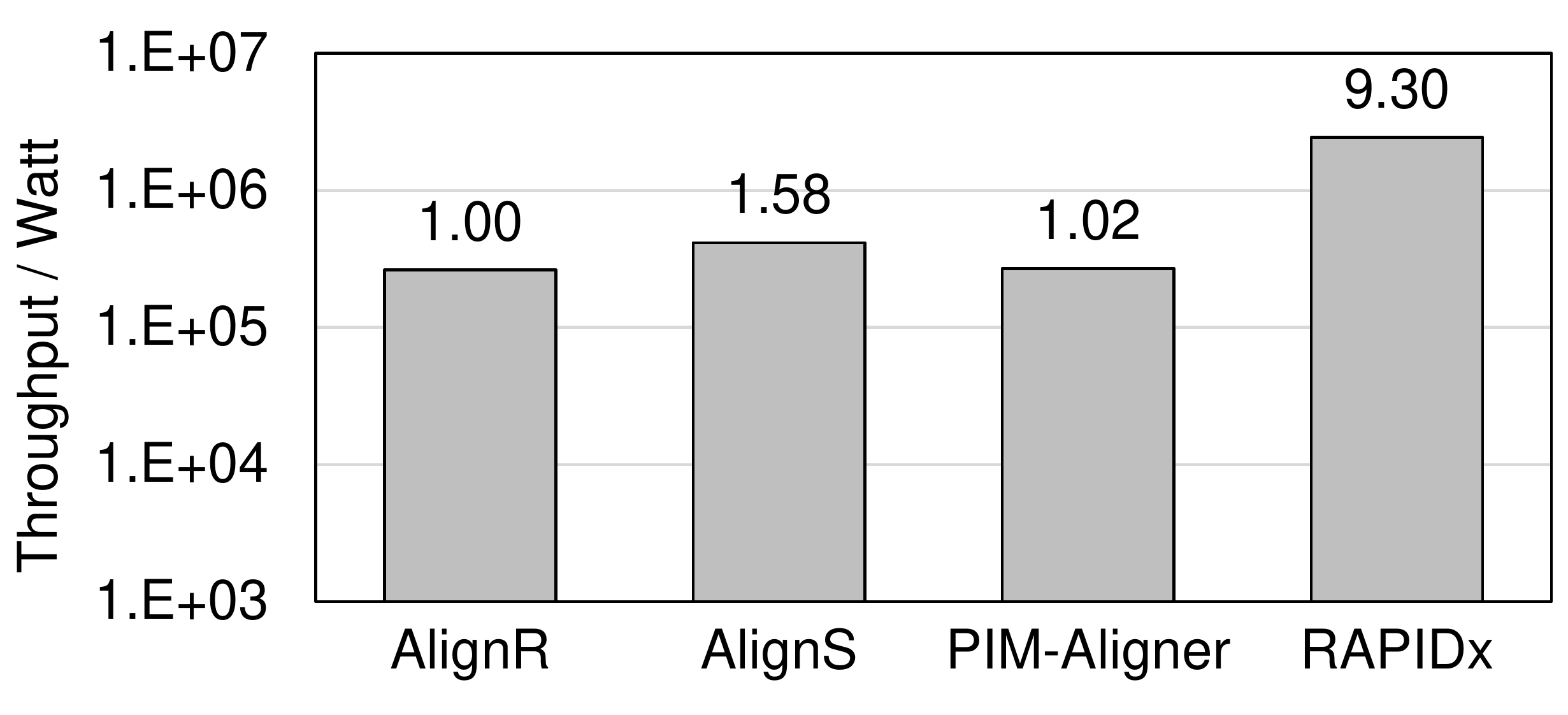}
		\caption{Energy efficiency comparison (in $\log$ scale) with PIM-based accelerators: AlignS \cite{angizi2019aligns}, AligneR \cite{zokaee2018aligner}, and PIM-Aligner \cite{angizi2020pim}.}
	\end{subfigure}
	\caption{Comparison with four PIM baselines, RAPID \cite{gupta2019rapid}, AlignS \cite{angizi2019aligns}, AligneR \cite{zokaee2018aligner}, and PIM-Aligner~\cite{angizi2020pim}.}
	\label{fig:pim_primitive}
\end{figure}

\subsubsection{Comparison with PIM Designs} \label{subsec:perf_comp_rapid}
Our previous work, RAPID \cite{gupta2019rapid}, is also a ReRAM-based PIM design for sequence alignment. First, we evaluate the reduction of processing latency and energy by adopting the parallelized DP alignment. The comparison of latency and energy with the original DP alignment for a single step of cells updating is shown in Fig. \ref{fig:pim_primitive} (a). RAPID uses the unoptimized DP alignment with 32-bit precision. The used PIM operations are the same as \Design. As a result, the parallelized DP alignment based on difference presentation yields $5.5\times$ latency reduction and $6.2\times$ energy reduction over the original DP alignment. The latency and energy consumed by forward DP computation are reduced by 82\% and 84\% over the previous RAPID, respectively. The gain comes from the reduced arithmetic precision from 32-bit to 5-bit as well as the parallelized computation. On the other hand, the reduction of latency and energy for traceback is less significant. Although the parallelized DP alignment requires less bit width, its traceback is more complicated and involves more computations than the original DP algorithm. The longest sequence support by \Design is 10kbp so we test the throughput of RAPID and \Design on this length in Fig. \ref{fig:pim_primitive} (a). \Design yields $9.7\times$ throughput improvement over RAPID due to the low complexity and high data parallelism provided by adaptive banded parallelized DP alignment.

In Fig. \ref{fig:pim_primitive} (b), we compare the energy efficiency with the other three PIM designs for short-read alignment, including AlignS \cite{angizi2019aligns}, AligneR \cite{zokaee2018aligner}, and PIM-Aligner~\cite{angizi2020pim}. The read length is 100bp and the alignment efficiency is measured by the alignment throughput (reads per second) divided by the power dissipation. \Design delivers $5.9\times$ to $9.3\times$ alignment efficiency compared to other PIM designs. It should be also noted that the area of mentioned PIM designs is: \Design (40.8mm$^2$), AlignR (36.1mm$^2$), AlignS (62.5mm$^2$), and PIM-Aligner (59.3mm$^2$). This shows that \Design achieves 8.4$\times$ to 13.3$\times$ throughput/W/mm$^2$ efficiency compared to other designs. This is because the optimized adaptive banded parallelized DP alignment in \Design significantly reduces computational complexity over the original full DP algorithm and allows to fully exploit the internal data parallelism of ReRAM. In comparison, AlignS, AligneR, and PIM-Aligner realize alignment based on FM-index algorithm, which requires multiple steps of computation and incurs data dependency \cite{huangfu2019medal}. %Besides, 
AlignS, AligneR, and PIM-Aligner only support fixed read length while \Design supports both short reads and long reads, making \Design  more scalable and reconfigurable.

\subsubsection{Performance Comparison on Short-read Alignment} \label{subsec:perf_comp_short}
\begin{figure}[t]
	\centering
	\includegraphics[width=0.9\linewidth]{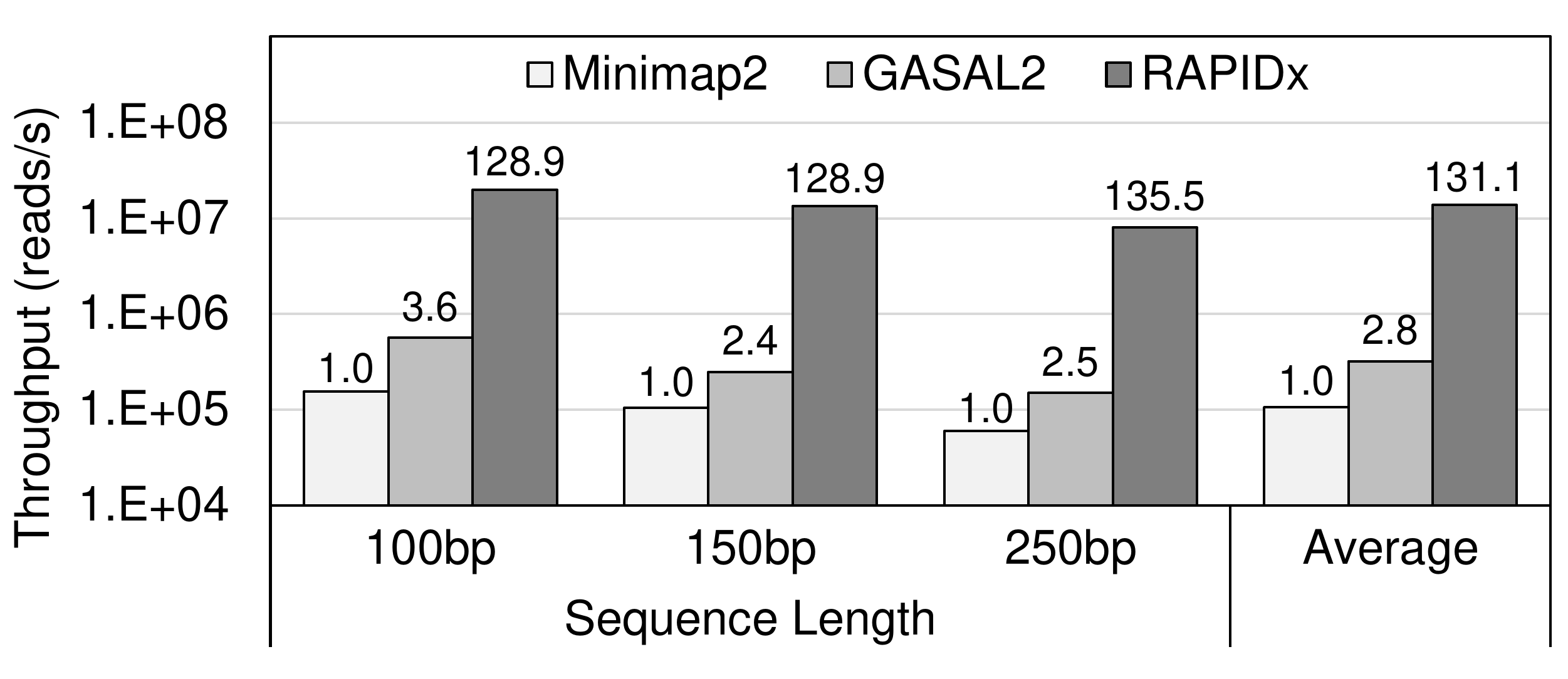}
	\caption{Alignment throughput comparison of \Design, GASAL2 \cite{gasal2}, and Minimap2 \cite{minimap2} for short reads.}
	\label{fig:cpu_gpu_comparsion}
\end{figure}

For alignment tasks on short reads, the length ranges from 100bp to 250bp and we use Minimap2 \cite{minimap2} as the CPU baseline and GASAL2 \cite{gasal2} as the GPU baseline. Fig. \ref{fig:cpu_gpu_comparsion} depicts the alignment throughput of \Design, Minimap2, and GASAL2 for short reads in $\log$ scale. The alignment throughputs for three tested accelerators slightly decrease as the sequence length grows. \Design on average delivers $131.1\times$ and $46.8\times$ throughput over Minimap2 and GASAL2, respectively. The processing latency of \Design is longer than the other two counterparts due to the fact that a single PIM operation of \Design requires longer latency than CPU and GPU. However, the row-parallel PIM operations provide higher computation parallelism. The proposed multi-level parallelism scheme ensures multiple reference and query sequences can be aligned in parallel, significantly increasing the data parallelism and PIM utilization. As a result, \Design achieves an average throughput of 13.9M reads/s for short-read alignment.

DP alignment is computation-intensive and the bottleneck of CPU is the limited computing cores. Even though GPU has much more computing capabilities than CPU, we observe that GASAL2 only yields $2.4\times$ to $3.6\times$ speedup over Minimap2 because Minimap2 uses a banded DP algorithm and multi-threading to reduce the complexity, thus improving the overall throughput. In comparison, GASAL2 requires more computing resources since it does not finely optimize the original DP alignment. \Design is an algorithm and hardware co-optimization that addresses the deficits of Minimap2 and GASAL2.

\subsubsection{Performance Comparison on Long-read Alignment}\label{subsec:perf_comp_long}
\begin{figure}[t]
	\centering
	\includegraphics[width=\columnwidth]{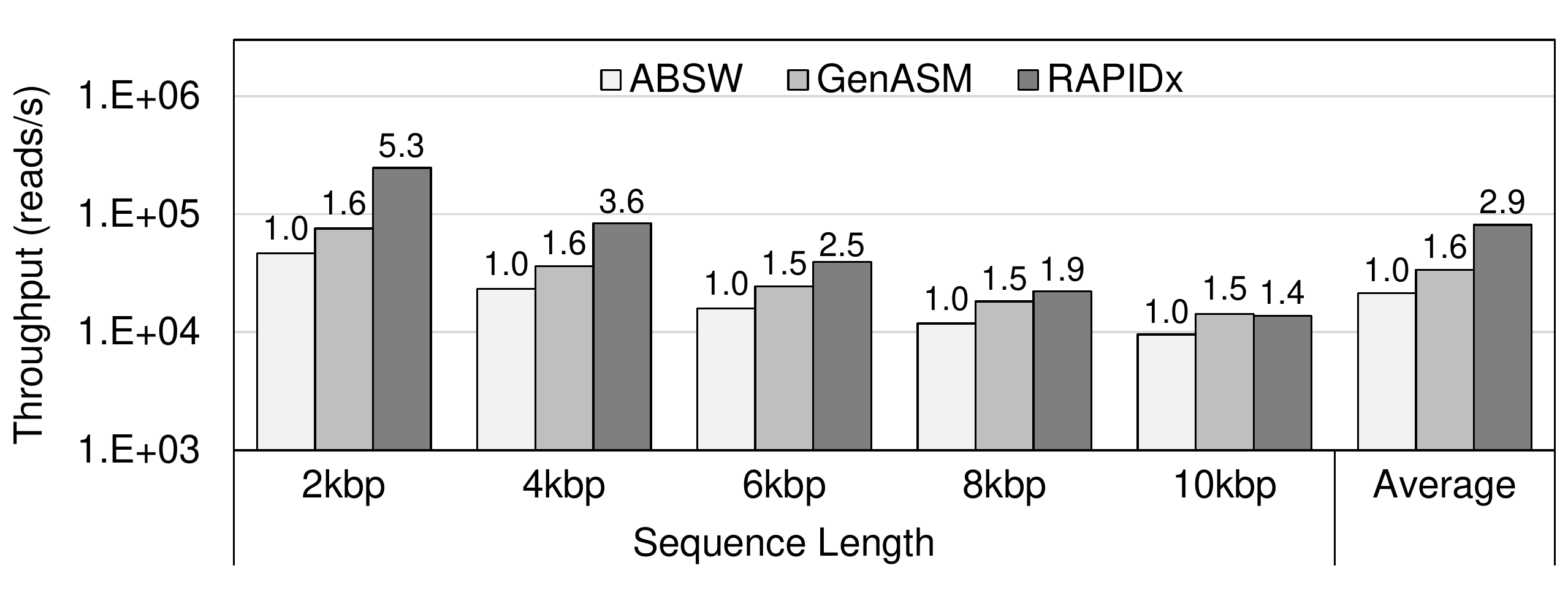}
	\caption{Alignment throughput comparison of GenASM \cite{genasm}, ABSW \cite{liao2018adaptively}, and \Design for long reads.}
	\label{fig:long_asic_comparsion}
\end{figure}

For long reads from 2kbp to 10kbp, ABSW \cite{liao2018adaptively} and GenASM \cite{genasm}, are adopted as the two ASIC baselines. The throughput comparison with ASIC for long-read alignment is shown in Fig. \ref{fig:long_asic_comparsion}, where the performance of ASIC baselines is scaled to 45nm process for the fair comparison. \Design achieves the highest throughput with an average speedup of $2.9\times$ and $1.8\times$ over ABSW and GenASM, respectively. Due to the limited on-chip memory space, both ABSW and GenASM are not able to store the entire traceback matrix for long reads. They rely on large off-chip memory to store the intermediate data. To realize alignment for long sequences, they use the overlapping scheme \cite{turakhia2018darwin} to divide the long sequence into short chunks and the neighbor chunks are overlapped. ABSW and GenASM need to consecutively process the short chunks. As a result, the overlapping area incurs additional computational complexity, which degrades the performance.

ABSW and \Design are based on banded DP algorithms. The difference between this work and ABSW is \Design adopts the optimized 5-bit parallelized DP alignment based on difference representations. ABSW uses 12-bit precision to ensure arithmetic precision for DP alignment. \Design's lower bit width reduces both the complexity and the memory footprint of DP alignment compared to ABSW. The other limitation of ABSW is it can only process a fixed bandwidth of 128 since a total of 128 processing elements (PEs) are implemented and dedicated to updating the wavefront of banded alignment. This means ABSW is only able to align one sequence at a time. In contrast, \Design accepts a batch of sequences and distributes them into different tiles to perform alignment in parallel.

\subsubsection{Performance Comparison on Edit Distance Computation} \label{subsec:perf_comp_edit}
\begin{figure}[t]
	\centering
	\includegraphics[width=\columnwidth]{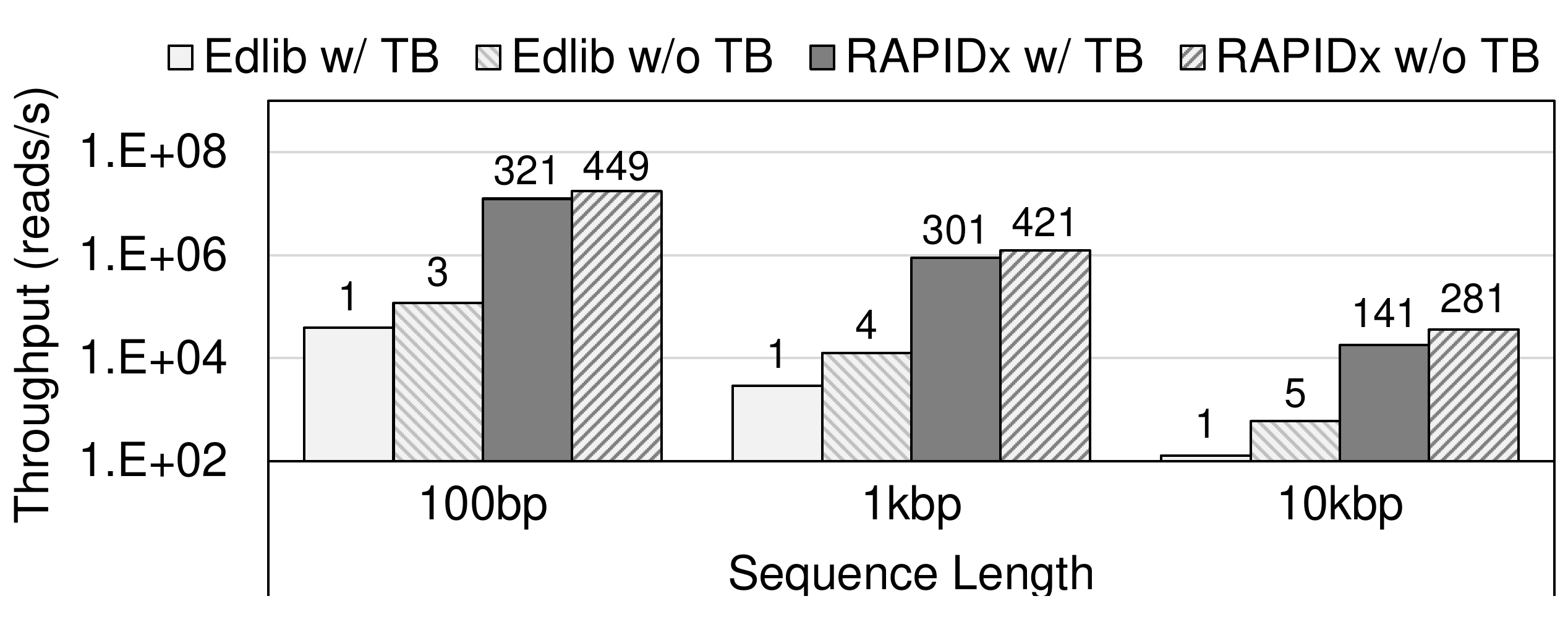}
	\caption{Throughput and latency comparison of \Design and Edlib \cite{vsovsic2017edlib} for edit distance computation.}
	\label{fig:edit_dist_comparison}
\end{figure}

To evaluate the performance of edit distance calculation, we compare \Design with Edlib \cite{vsovsic2017edlib} on three lengths (100bp, 1kbp, and 10kbp). Fig. \ref{fig:edit_dist_comparison} shows the throughput of \Design and Edlib with or without traceback process. Knowing the edit distance of two sequences is enough for some scenarios, without the need for traceback process. So we test the cases with or without traceback. The throughput of \Design with traceback is $141\times$ to $321\times$ over Edlib. After disabling the traceback, the speedup of \Design is less significant. $56\times$ to $149\times$ improvements of \Design are observed compared to Edlib. Although Edlib adopts optimized Myers's bit-vector algorithm \cite{myers1999fast} with banded alignment to increase computation efficient, it is a single-thread program only able to access limited computing resources of CPU. Hence, the performance dramatically decreases after enabling traceback.

\subsection{Discussions}
\textbf{Host-\Design System Design:}
\Design is a PIM-based domain-specific accelerator and works as the domain-specific co-processor for speeding up computation-intensive genome sequence alignments. We consider a system that transfers data between \Design and the host. The sequencing and configuration data are sent from the host to \Design. We estimate the memory bandwidth required by \Design and the results show that required memory bandwidth decreases when sequence length grows. The required peak memory bandwidth is 1.41GB/s at 100bp.
%, which matches the maximum read/write bandwidth of ReRAM subarray that uses 64-bit column width. 
For the host side, the popular DDR4 Dual-Inline Memory Module (DIMM) that provides over 12.8GB/s data rate can easily satisfy the bandwidth requirement. The other consideration is the processing latency. As pointed out in Section \ref{subsec:perf_eval}, \Design requires longer latency than CPU. Considering that genome sequence alignment is not a latency-sensitive task, the long latency will not become a major factor that limits system performance. Hence, \Design can be integrated into existing computer machines with negligible hardware modifications.

\noindent
\textbf{Flexible Scoring Functions:} The affine gap penalty of DP alignment will be changed according to different application scenarios. \Design is able to flexibly support various scoring functions. When the gap open penalty $o$ equals the gap extension penalty $e$, the affine gap penalty becomes a linear gap penalty scoring. If $e=0$, \Design implements a constant gap penalty where only opening a gap leads to a penalty, discouraging the number of gaps but tends to result in long gaps. Whereas, if $o \neq e$ and both of $o$ and $e$ are non-zero values, we have affine gap penalty, which is the widely used gap penalty model for DNA alignment. The affine gap penalty tries to align the given sequences with fewer and smaller gaps as compared to the constant gap penalty. No architectural and data flow modifications need to be made to \Design if we want to switch between different scoring functions. The support for flexible scoring is realized by setting associated constant values into the intermediate data rows of CM before alignment.

\noindent
\textbf{ReRAM's Write Endurance:} ReRAM cell has limited write endurance, so \Design will fail after exceeding the endurance limit. As shown in Fig. \ref{fig:data_flow} (c), the wavefront alignment at each iteration needs to write the rows in the computing region once. Fig. \ref{fig:adaptive_band} shows the required number of iterations equals to the sum of reference and query sequences' lengths. We can apply wear leveling techniques to reduce the imbalance effect, thus extending the write endurance of ReRAM. The wear leveling is realized via moving the computing region over the row dimension. Specifically, this can be done through changing the writing address without additional overhead. Moreover, we observe some ReRAM devices \cite{luo2019nb} provide $10^{12}$ write endurance. In this case, \Design can align over $10^{14}$ sequences with length 150bp. We notice that one of the most advanced next-generation sequencing (NGS) platforms from Illumina, \textit{NextSeq 1000 \& 2000}, generates a maximum 1.2 billion reads (each has a length of 150bp) in 11 to 48 hours \cite{illumina_ngs}. Therefore, each \Design is able to support the alignment task of each NGS sequencer for at least 100 years.

%%%%%%%%%%%%%%%%%%%%%%%%%%%%%%%%%%%%%%%%%%%%%%%%%%%%%%%%%%%%%%
\section{Conclusion}
In this work, we propose a novel PIM accelerator, \Design, for sequence alignment. We leverage the parallelized DP algorithm using difference representation to reduce the required data width from 32-bit to 5-bit integers. Based on this, we propose adaptive banded parallelized DP alignment to adaptively adjust the bandwidth and wavefront direction, reducing the quadratic complexity to near-linear complexity while only incurring 0.15\% accuracy degradation. 
Then we present the PIM architecture on ReRAM that exploits four-level data parallelism to efficiently implement the proposed algorithm. We develop peripheral circuits and row-parallel PIM data flow to support in-situ alignment with low latency. The evaluation results demonstrate that \Design provides $131.1\times$ and $46.8\times$ better short-read alignment throughput compared to CPU and GPU baselines, respectively. For long-read alignment, \Design delivers up to $2.9\times$ and $9.3\times$ throughput improvements compared to state-of-the-art ASIC and PIM accelerators.

% \appendices

\section*{Acknowledgment}
This work was supported in part by CRISP, one of six centers in JUMP (an SRC program sponsored by DARPA), SRC Global Research Collaboration (GRC) grant, and NSF grants \#1826967, \#1911095, \#2003279, \#2112665, \#2112167, and \#2100237.

The authors thank the anonymous reviewers from JETC and the constructive discussion with Behnam Khaleghi to help improve the quality of this work.



{
    \scriptsize
    \bibliographystyle{IEEEtran}
    \bibliography{IEEEabrv,mybib}

\begin{thebibliography}{10}
\providecommand{\url}[1]{#1}
\csname url@samestyle\endcsname
\providecommand{\newblock}{\relax}
\providecommand{\bibinfo}[2]{#2}
\providecommand{\BIBentrySTDinterwordspacing}{\spaceskip=0pt\relax}
\providecommand{\BIBentryALTinterwordstretchfactor}{4}
\providecommand{\BIBentryALTinterwordspacing}{\spaceskip=\fontdimen2\font plus
\BIBentryALTinterwordstretchfactor\fontdimen3\font minus
  \fontdimen4\font\relax}
\providecommand{\BIBforeignlanguage}[2]{{%
\expandafter\ifx\csname l@#1\endcsname\relax
\typeout{** WARNING: IEEEtran.bst: No hyphenation pattern has been}%
\typeout{** loaded for the language `#1'. Using the pattern for}%
\typeout{** the default language instead.}%
\else
\language=\csname l@#1\endcsname
\fi
#2}}
\providecommand{\BIBdecl}{\relax}
\BIBdecl

\bibitem{seqoverview}
F.~E. Dewey \emph{et~al.}, ``Dna sequencing: clinical applications of new dna
  sequencing technologies,'' \emph{Circulation}, vol. 125, no.~7, pp. 931--944,
  2012.

\bibitem{beast}
A.~J. Drummond and A.~Rambaut, ``Beast: Bayesian evolutionary analysis by
  sampling trees,'' \emph{Evolutionary Biology}, vol.~7, no.~1, pp. 1--8, 2007.

\bibitem{minoritydetection}
S.~J. Watson \emph{et~al.}, ``Viral population analysis and minority-variant
  detection using short read next-generation sequencing,'' \emph{Philosophical
  Transactions of the Royal Society B: Biological Sciences}, vol. 368, no.
  1614, p. 20120205, 2013.

\bibitem{nw}
S.~B. Needleman and C.~D. Wunsch, ``A general method applicable to the search
  for similarities in the amino acid sequence of two proteins,'' \emph{JMB},
  vol.~48, no.~3, pp. 443--453, 1970.

\bibitem{smith1981identification}
T.~F. Smith \emph{et~al.}, ``Identification of common molecular subsequences,''
  \emph{JMB}, vol. 147, no.~1, pp. 195--197, 1981.

\bibitem{vsovsic2017edlib}
M.~{\v{S}}o{\v{s}}i{\'c} and M.~{\v{S}}iki{\'c}, ``Edlib: a c/c++ library for
  fast, exact sequence alignment using edit distance,'' \emph{Bioinformatics},
  vol.~33, no.~9, pp. 1394--1395, 2017.

\bibitem{minimap2}
H.~Li, ``Minimap2: pairwise alignment for nucleotide sequences,''
  \emph{Bioinformatics}, vol.~34, no.~18, pp. 3094--3100, 2018.

\bibitem{gasal2}
N.~Ahmed \emph{et~al.}, ``Gasal2: a gpu accelerated sequence alignment library
  for high-throughput ngs data,'' \emph{Bioinformatics}, vol.~20, no.~1, pp.
  1--20, 2019.

\bibitem{bwamem}
H.~Li and R.~Durbin, ``Fast and accurate long-read alignment with
  burrows--wheeler transform,'' \emph{Bioinformatics}, vol.~26, no.~5, pp.
  589--595, 2010.

\bibitem{langmead2012fast}
B.~Langmead and S.~L. Salzberg, ``Fast gapped-read alignment with bowtie 2,''
  \emph{Nature methods}, vol.~9, no.~4, pp. 357--359, 2012.

\bibitem{liao2018adaptively}
Y.-L. Liao \emph{et~al.}, ``Adaptively banded smith-waterman algorithm for long
  reads and its hardware accelerator,'' in \emph{International Conference on
  Application-specific Systems, Architectures and Processors}, 2018, pp. 1--9.

\bibitem{genasm}
D.~S. Cali \emph{et~al.}, ``Genasm: A high-performance, low-power approximate
  string matching acceleration framework for genome sequence analysis,'' in
  \emph{IEEE/ACM MICRO}, 2020, pp. 951--966.

\bibitem{turakhia2018darwin}
Y.~Turakhia \emph{et~al.}, ``Darwin: A genomics co-processor provides up to
  15,000$\times$ acceleration on long read assembly,'' in \emph{ASPLOS}, 2018.

\bibitem{de2016cudalign}
E.~F. de~Oliveira~Sandes \emph{et~al.}, ``Cudalign 4.0: Incremental speculative
  traceback for exact chromosome-wide alignment in gpu clusters,'' \emph{IEEE
  Transactions on Parallel and Distributed Systems}, vol.~27, no.~10, pp.
  2838--2850, 2016.

\bibitem{arram2017leveraging}
J.~Arram \emph{et~al.}, ``Leveraging fpgas for accelerating short read
  alignment,'' \emph{IEEE/ACM TCBB}, vol.~14, no.~3, pp. 668--677, 2017.

\bibitem{genome_cost}
``Dna sequencing costs: Data from the nhgri genome sequencing program (gsp),''
  \url{www.genome.gov/sequencingcostsdata}.

\bibitem{genbank}
``Genbank and wgs statistics,''
  \seqsplit{\url{https://www.ncbi.nlm.nih.gov/genbank/statistics/}}.

\bibitem{wenger2019accurate}
A.~M. Wenger \emph{et~al.}, ``Accurate circular consensus long-read sequencing
  improves variant detection and assembly of a human genome,'' \emph{Nature
  biotechnology}, vol.~37, no.~10, pp. 1155--1162, 2019.

\bibitem{gokhale1995processing}
M.~Gokhale \emph{et~al.}, ``Processing in memory: The terasys massively
  parallel pim array,'' \emph{Computer}, vol.~28, no.~4, pp. 23--31, 1995.

\bibitem{ahn2015pim}
J.~Ahn \emph{et~al.}, ``Pim-enabled instructions: a low-overhead,
  locality-aware processing-in-memory architecture,'' in \emph{ISCA}, 2015, pp.
  336--348.

\bibitem{li2016pinatubo}
S.~Li \emph{et~al.}, ``Pinatubo: a processing-in-memory architecture for bulk
  bitwise operations in emerging non-volatile memories,'' in \emph{DAC}, 2016,
  p. 173.

\bibitem{gupta2019nnpim}
S.~Gupta \emph{et~al.}, ``Nnpim: A processing in-memory architecture for neural
  network acceleration,'' \emph{IEEE Transactions on Computers}, vol.~68,
  no.~9, pp. 1325--1337, 2019.

\bibitem{kaplan2017resistive}
R.~Kaplan \emph{et~al.}, ``A resistive cam processing-in-storage architecture
  for dna sequence alignment,'' \emph{IEEE Micro}, vol.~37, no.~4, pp. 20--28,
  2017.

\bibitem{bioseal}
------, ``Bioseal: In-memory biological sequence alignment accelerator for
  large-scale genomic data,'' in \emph{ACM International Systems and Storage
  Conference}, 2020, pp. 36--48.

\bibitem{huangfu2018radar}
W.~Huangfu, S.~Li, X.~Hu, and Y.~Xie, ``Radar: a 3d-reram based dna alignment
  accelerator architecture,'' in \emph{DAC}, 2018, pp. 1--6.

\bibitem{angizi2019aligns}
S.~Angizi \emph{et~al.}, ``Aligns: A processing-in-memory accelerator for dna
  short read alignment leveraging sot-mram,'' in \emph{DAC}, 2019, pp. 1--6.

\bibitem{zokaee2018aligner}
F.~Zokaee \emph{et~al.}, ``Aligner: A process-in-memory architecture for short
  read alignment in rerams,'' \emph{IEEE Computer Architecture Letters},
  vol.~17, no.~2, pp. 237--240, 2018.

\bibitem{gupta2019rapid}
S.~Gupta \emph{et~al.}, ``Rapid: A reram processing in-memory architecture for
  dna sequence alignment,'' in \emph{IEEE/ACM ISLPED}, 2019, pp. 1--6.

\bibitem{liu2008barking}
K.~Liu \emph{et~al.}, ``Barking up the wrong treelength: the impact of gap
  penalty on alignment and tree accuracy,'' \emph{IEEE/ACM Transactions on
  Computational Biology and Bioinformatics}, vol.~6, no.~1, pp. 7--21, 2008.

\bibitem{seedex}
D.~Fujiki \emph{et~al.}, ``Seedex: A genome sequencing accelerator for optimal
  alignments in subminimal space,'' in \emph{IEEE/ACM MICRO}, 2020, pp.
  937--950.

\bibitem{chao1992aligning}
K.-M. Chao \emph{et~al.}, ``Aligning two sequences within a specified diagonal
  band,'' \emph{Bioinformatics}, vol.~8, no.~5, pp. 481--487, 1992.

\bibitem{gupta2018felix}
S.~Gupta \emph{et~al.}, ``Felix: Fast and energy-efficient logic in memory,''
  in \emph{IEEE/ACM ICCAD}, 2018, pp. 1--7.

\bibitem{bwt}
M.~Burrows and D.~Wheeler, ``A block-sorting lossless data compression
  algorithm,'' in \emph{Digital SRC Research Report}, 1994.

\bibitem{myers1999fast}
G.~Myers, ``A fast bit-vector algorithm for approximate string matching based
  on dynamic programming,'' \emph{Journal of the ACM (JACM)}, vol.~46, no.~3,
  pp. 395--415, 1999.

\bibitem{suzuki2018introducing}
H.~Suzuki and M.~Kasahara, ``Introducing difference recurrence relations for
  faster semi-global alignment of long sequences,'' \emph{Bioinformatics},
  vol.~19, no.~1, pp. 33--47, 2018.

\bibitem{angizi2020pim}
S.~Angizi \emph{et~al.}, ``Pim-aligner: A processing-in-mram platform for
  biological sequence alignment,'' in \emph{DATE}, 2020, pp. 1265--1270.

\bibitem{altschul1990basic}
S.~F. Altschul \emph{et~al.}, ``Basic local alignment search tool,''
  \emph{Journal of Molecular Biology}, vol. 215, no.~3, pp. 403--410, 1990.

\bibitem{lipman1985rapid}
D.~J. Lipman and W.~R. Pearson, ``Rapid and sensitive protein similarity
  searches,'' \emph{Science}, vol. 227, no. 4693, pp. 1435--1441, 1985.

\bibitem{banerjee2019asap}
S.~S. Banerjee \emph{et~al.}, ``Asap: Accelerated short-read alignment on
  programmable hardware,'' \emph{IEEE Transactions on Computers}, 2019.

\bibitem{gotoh1982improved}
O.~Gotoh, ``An improved algorithm for matching biological sequences,''
  \emph{Journal of molecular biology}, vol. 162, no.~3, pp. 705--708, 1982.

\bibitem{lee2020bit}
K.~Lee \emph{et~al.}, ``Bit parallel 6t sram in-memory computing with
  reconfigurable bit-precision,'' in \emph{DAC}, 2020, pp. 1--6.

\bibitem{boukhobza2017emerging}
J.~Boukhobza \emph{et~al.}, ``Emerging nvm: A survey on architectural
  integration and research challenges,'' \emph{ACM Transactions on Design
  Automation of Electronic Systems}, vol.~23, no.~2, pp. 1--32, 2017.

\bibitem{reis2018computing}
D.~Reis \emph{et~al.}, ``Computing in memory with fefets,'' in \emph{ISLPED},
  2018, pp. 1--6.

\bibitem{kim2021embedded}
M.~Kim \emph{et~al.}, ``An embedded nand flash-based compute-in-memory array
  demonstrated in a standard logic process,'' \emph{IEEE Journal of Solid-State
  Circuits}, vol.~57, no.~2, pp. 625--638, 2021.

\bibitem{xue202116}
C.-X. Xue \emph{et~al.}, ``A 22nm 4mb 8b-precision reram computing-in-memory
  macro with 11.91 to 195.7 tops/w for tiny ai edge devices,'' in \emph{ISSCC},
  vol.~64, 2021, pp. 245--247.

\bibitem{talati2016logic}
N.~Talati \emph{et~al.}, ``Logic design within memristive memories using
  memristor-aided logic (magic),'' \emph{IEEE Transactions on Nanotechnology},
  vol.~15, no.~4, pp. 635--650, 2016.

\bibitem{borghetti2010memristive}
J.~Borghetti \emph{et~al.}, ``Memristive switches enable stateful logic
  operations via material implication,'' \emph{Nature}, vol. 464, no. 7290, pp.
  873--876, 2010.

\bibitem{jang2018memristive}
B.~C. Jang \emph{et~al.}, ``Memristive logic-in-memory integrated circuits for
  energy-efficient flexible electronics,'' \emph{Advanced Functional
  Materials}, vol.~28, no.~2, p. 1704725, 2018.

\bibitem{kvatinsky2015vteam}
S.~Kvatinsky \emph{et~al.}, ``Vteam: a general model for voltage-controlled
  memristors,'' \emph{IEEE TCAS II}, vol.~62, no.~8, pp. 786--790, 2015.

\bibitem{haj2018efficient}
A.~Haj-Ali \emph{et~al.}, ``Efficient algorithms for in-memory fixed point
  multiplication using magic,'' in \emph{IEEE ISCAS}, 2018, pp. 1--5.

\bibitem{kvatinsky2014magic}
S.~Kvatinsky \emph{et~al.}, ``{M}{A}{G}{I}{C} -- memristor-aided logic,''
  \emph{TCAS II}, vol.~61, no.~11, 2014.

\bibitem{imani2019floatpim}
M.~Imani \emph{et~al.}, ``Floatpim: In-memory acceleration of deep neural
  network training with high precision,'' in \emph{ISCA}, 2019.

\bibitem{suzuki2017acceleration}
H.~Suzuki and M.~Kasahara, ``Acceleration of nucleotide semi-global alignment
  with adaptive banded dynamic programming,'' \emph{BioRxiv}, p. 130633, 2017.

\bibitem{yang2013memristive}
J.~J. Yang \emph{et~al.}, ``Memristive devices for computing,'' \emph{Nature
  nanotechnology}, vol.~8, no.~1, pp. 13--24, 2013.

\bibitem{dong2012nvsim}
X.~Dong \emph{et~al.}, ``Nvsim: A circuit-level performance, energy, and area
  model for emerging nonvolatile memory,'' \emph{IEEE TCAD}, vol.~31, no.~7,
  pp. 994--1007, 2012.

\bibitem{stine2007freepdk}
J.~E. Stine \emph{et~al.}, ``Freepdk: An open-source variation-aware design
  kit,'' in \emph{IEEE International Conference on Microelectronic Systems
  Education}, 2007, pp. 173--174.

\bibitem{cacti}
N.~Muralimanohar \emph{et~al.}, ``Cacti 6.0: A tool to model large caches,''
  \emph{HP laboratories}, vol.~27, p.~28, 2009.

\bibitem{GRCh38}
N.~C. for Biotechnology~Information, ``Genome reference consortium human build
  38,'' \url{https://www.ncbi.nlm.nih.gov/assembly/GCF\_000001405.26}, 2013.

\bibitem{ono2013pbsim}
Y.~Ono \emph{et~al.}, ``Pbsim: Pacbio reads simulator—toward accurate genome
  assembly,'' \emph{Bioinformatics}, vol.~29, no.~1, pp. 119--121, 2013.

\bibitem{mason}
M.~Holtgrewe, ``Mason--a read simulator for second generation sequencing
  data,'' \emph{Technical Report FU Berlin}, 2010.

\bibitem{huangfu2019medal}
W.~Huangfu \emph{et~al.}, ``Medal: Scalable dimm based near data processing
  accelerator for dna seeding algorithm,'' in \emph{IEEE/ACM MICRO}, 2019, pp.
  587--599.

\bibitem{luo2019nb}
Q.~Luo \emph{et~al.}, ``Nb$_{1-x}$ o2 based universal selector with ultra-high
  endurance ($>10^{12}$), high speed (10ns) and excellent v$_{th}$ stability,''
  in \emph{Symposium on VLSI Technology}, 2019, pp. T236--T237.

\bibitem{illumina_ngs}
``Illumina sequencing platforms,''
  \seqsplit{https://www.illumina.com/systems/sequencing-platforms.html}.

\end{thebibliography}
}


% \end{IEEEbiography}


\clearpage

\begin{IEEEbiographynophoto}{Weihong Xu}
received the B.E. degree in information
engineering and M.E. degree information science and engineering from Southeast University, Nanjing, China, in 2017 and 2020, respectively. 
He is currently pursing the Ph.D. degree with the Department of Computer Science and Engineering, University of California at San Diego, La Jolla, CA, USA.

He is a member of the System Energy Efficiency Laboratory, University of California at San Diego, where he works on emerging memory, storage and domain-specific acceleration systems.
His current research interests include algorithm and hardware co-design based on near-data processing for deep learning, database, and bioinformatics applications.

\end{IEEEbiographynophoto}

\begin{IEEEbiographynophoto}{Saransh Gupta}
received his Ph.D. degree from the University of California, San Diego, La Jolla, CA, USA in 2021. He received his B.E. (Hons) in Electrical and Electronics Engineering from Birla Institute of Technology \& Science, Pilani - K.K. Birla Goa Campus in 2016 and M.S. in Electrical and Computer Engineering from University of California San Diego in 2018. 
His research interests include circuit, architecture, and system level aspects of emerging computing paradigms.
\end{IEEEbiographynophoto}

\begin{IEEEbiographynophoto}{Niema Moshiri}
received his Ph.D. degree from the University of California, San Diego, La Jolla, CA, USA in 2019. He is an Assistant Teaching Professor in the Computer Science \& Engineering Department at the University of California, San Diego. He works on computational biology, with a research focus on viral phylogenetics and epidemiology. He also places a heavy emphasis on teaching, namely on the development of online educational content, primarily Massive Adaptive Interactive Texts (MAITs).

\end{IEEEbiographynophoto}

\begin{IEEEbiographynophoto}{Tajana Šimunić Rosing (Fellow, IEEE)}
received her Ph.D. degree from Stanford University, Stanford, CA, USA, in 2001. She is a Professor, a Holder of the Fratamico Endowed Chair, and the Director of System Energy Efficiency Laboratory, University of California at San Diego, La Jolla, CA, USA. From 1998 to 2005, she was a full-time Research Scientist with HP Labs, Palo Alto, CA, USA, while also leading research efforts with Stanford University, Stanford, CA, USA. She was a Senior Design Engineer with Altera Corporation, San Jose, CA, USA. She is leading a number of projects, including efforts funded by DARPA/SRC JUMP CRISP program with focus on design of accelerators for analysis of big data, DARPA and NSF funded projects on hyperdimensional computing, and SRC funded project on IoT system reliability and maintainability. Her current research interests include energy-efficient computing, cyber–physical, and distributed systems.
\end{IEEEbiographynophoto}


%\vfill


\end{document}